\documentclass[11pt]{article}
\renewcommand{\theequation}{\thesection.\arabic{equation}}

\usepackage{amsfonts}
\usepackage{amsbsy}
\usepackage{amssymb}

\usepackage{dcolumn}
\newcolumntype{d}[1]{D{.}{\cdot}{#1}}

\usepackage[pdftex]{graphicx}

\newcommand{\BE}{\begin{equation}}
\newcommand{\EE}{\end{equation}}
\newcommand{\BA}{\begin{eqnarray}}
\newcommand{\EA}{\end{eqnarray}}
\newcommand{\half}{{\textstyle{\frac{1}{2}}}}

\newcommand{\quarter}{{\textstyle{\frac{1}{4}}}}
\newcommand{\Pp}{{\scriptstyle{{\rm P}}}}
\newcommand{\EC}{{\scriptscriptstyle \rm EC}}
\newcommand{\msbar}{\overline{\rm MS}}

\newcommand{\kplus}{(k+1)^{th}}

\newcommand{\adot}{a^\star}
\newcommand{\ap}{a_{\rm p}}
\newcommand{\udot}{u^\star}
\newcommand{\up}{u_{\rm p}}

\begin{document}

\thispagestyle{empty}


\vspace*{13mm}
\begin{center}
              {\LARGE\bf  Exploring arbitrarily high orders  \\ 
              \vspace*{1mm} of optimized perturbation theory \\ \vspace*{4.5mm} 
              in QCD with $\boldsymbol{n_f} \, {\mathbf{\to 16\half}}$} 
\vspace{20mm}\\
{\large P. M. Stevenson}
\vspace{4mm}\\
{\it
T.W. Bonner Laboratory, Department of Physics and Astronomy,\\
Rice University, Houston, TX 77251, USA}

\vspace{22mm}

{\bf Abstract:}

\end{center}

\begin{quote}
Perturbative QCD with $n_f$ flavours of massless quarks becomes simple in the hypothetical 
limit $n_f \to 16\half$, where the leading $\beta$-function coefficient vanishes.  
The Banks-Zaks (BZ) expansion in 
$a_0 \equiv \scriptstyle{\frac{8}{321}}(16{\scriptstyle{\frac{1}{2}}}-n_f)$ is 
straightforward to obtain from perturbative results in  $\msbar$ or any renormalization 
scheme (RS) whose $n_f$ dependence is `regular.'  However, `irregular' RS's are 
perfectly permissible and should ultimately lead to the same BZ results.  We show here that 
the `optimal' RS determined by the Principle of Minimal Sensitivity does yield the same 
BZ-expansion results when all orders of perturbation theory are taken into account.  
The BZ limit provides an arena for exploring optimized perturbation theory at arbitrarily 
high orders.  These explorations are facilitated by a `master equation' expressing the 
optimization conditions in the fixed-point limit.  We find an intriguing strong/weak coupling 
duality $a \to  {a^*}^2/a$ about the fixed point $a^*$.  

\end{quote}

\newpage

\section{Introduction}
\setcounter{equation}{0}

      The initial impulse for these investigations was a concern with the compatibility of 
the Banks-Zaks (BZ) expansion \cite{BZ}-\cite{BZlett} with renormalization-scheme (RS) 
invariance \cite{RG}.  In dimensional regularization the $\beta$ function naturally has a 
term $-\epsilon a$ which strongly affects any zero near the origin.  Can one safely 
take $\epsilon \to 0$ first and {\it then} take $n_f \to 16 \half$, or do these limits somehow clash?  
Our results here basically resolve those concerns;  the BZ expansion appears to be fully 
compatible with RS invariance in the sense that ``optimized perturbation theory'' (OPT) 
\cite{OPT}, which enforces local RS invariance in each order,  ultimately yields the same 
BZ results.  

       The BZ expansion is normally discussed only within a restricted class of `regular' 
schemes.  However, infinitely many schemes -- and in some sense most schemes -- 
are not `regular.'   In particular, the ``optimal'' scheme is not.  In `regular' schemes 
one needs only $k$ terms of the perturbation series to obtain $k$ terms of the BZ 
expansion, but in other schemes the information needed is distributed among higher-order 
terms \cite{Caveny}.  In general all orders are required.  Turning that observation around, 
the BZ expansion can be viewed as a ``playground'' in which one can analytically investigate 
arbitrarily high orders of OPT in QCD.  Admittedly, this adopts the 
``drunk-under-the-lamppost'' principle of looking, not where we really want to, but where 
there is enough light to make a search.  The deep and difficult issues that we would like to 
study -- ``renormalons'' and factorially growing coefficients -- are simply absent in the 
BZ limit.  Nevertheless, we believe our search provides some interesting insights 
and employs some methods that may have wider applicability.  

      Infrared fixed points and divergent perturbation series were no part 
of the motivation for OPT \cite{OPT}, but OPT has important consequences for both 
these topics.  Fixed points in OPT are discussed in Refs.~\cite{KSS}-\cite{unfixed}.  
Such infrared behaviour was found for $R_{e^+e^-}$ at third order for all $n_f$ 
\cite{CKL,lowen}, though error estimates at low $n_f$ are large.\footnote{
Also, other physical quantities behave rather differently \cite{GardiK}.
The idea \cite{BZlett,Caveny} that the BZ expansion can be extrapolated, crudely, to 
low $n_f$ no longer seems tenable \cite{unfixed}.  The ``freezing'' behaviour at small 
$n_f$, confirmed at fourth order \cite{OPTnew}-\cite{unfixed}, seems instead to stem 
from somewhat different physics.
}

     The role of OPT in taming high-order perturbation theory was investigated in 
Ref.~\cite{optult}.  A toy example, involving an alternating factorial series, showed 
that even when the perturbation series is badly divergent in any fixed RS, the sequence 
of optimized approximants can converge.  This ``induced convergence'' mechanism 
(related to the idea of ``order-dependent mappings'' \cite{ZinnJ}) has been shown to 
operate \cite{IndConvAHO} in the anharmonic oscillator and $\phi^4$ field theories in the 
variational perturbation theory of Refs.~\cite{CK}-\cite{deltaexp}.  In QCD ``induced 
convergence'' of OPT has been investigated in the large-$b$ approximation \cite{Acoleyen}.  
It has also been shown \cite{Beneke} that adjusting the renormalization scale with increasing 
order --- which happens naturally in OPT \cite{optult} --- can indeed have dramatic and 
beneficial effects on series behaviour.  In the present paper we work in the small-$b$ 
approximation (the BZ limit), where the issues are rather different.  In particular, the role 
of optimizing other aspects of the RS, besides the renormalization scale, come to the fore.  

The plan of the paper is as follows.  Following some preliminaries in Sect.~2, the BZ expansion, 
as obtained from `regular' schemes, is summarized in Sect.~3, and we note that it is suffices to
consider two infrared quantities, ${\cal R}^*$ and $\gamma^*$.  Sect.~4 briefly reviews 
OPT\@.  Sect.~5 presents OPT results in the BZ limit, up to $19^{th}$ order.  Sect.~6 describes 
analytic methods for studying OPT at arbitrarily high orders.  It also introduces a crude 
approximation, ``NLS,'' and a better approximation, ``PWMR\@.''  These approximations, 
applied to the BZ limit, are explored in detail in Sects. 7 and 8.  From these results 
we see that OPT, taken to all orders, does reproduce the expected BZ-limit results, and we gain 
some insight into how OPT's subtle features conspire to produce accurate results 
and a rather well-behaved series for ${\cal R}^*$.  In Sect.~9 we show that all-orders OPT 
reproduces higher terms in the BZ expansion correctly, and in Sect.~10 we point out an intriguing 
$a \to {a^*}^2/a$ duality.  Our conclusions are summarized in Sect. 11.  (Two appendices 
discuss (a) some subtleties associated with the critical exponent $\gamma^*$ 
\cite{Gross}-\cite{effexp} and (b) the pinch mechanism \cite{unfixed}, which is a way that 
a finite infrared limit can occur in OPT without a fixed point.  This mechanism is 
probably not directly relevant in the BZ limit, though it nearly is.)  

\section{Preliminaries}
\setcounter{equation}{0}

     Consider a suitably normalized, perturbatively calculable, physical quantity ${\cal R}$ 
with a perturbation series 
\BE 
{\cal R} = a (1+r_1 a+ r_2 a^2 + r_3 a^3 + \ldots ) ,
\EE
where $a \equiv \alpha_s/\pi$ is the couplant of some particular renormalization scheme (RS).    
(More generally ${\cal R}$ can start $a^\Pp (1+ \ldots)$ but in this paper we will consider 
only $\Pp=1$.)  
The physical quantity ${\cal R}$ could be a function of several experimentally 
defined parameters.  One may always single out one parameter, ``$Q$,'' with 
dimensions of energy and let all other parameters be dimensionless.  (The precise definition of 
$Q$ in any specific case may be left to the reader; it is needed only to explain which quantities 
are, or are not, $Q$ dependent.)  For dimensional reasons the $r_i$ can depend on $Q$ and the 
renormalization scale $\mu$ only through the ratio $\mu/Q$. 

The physical quantity ${\cal R}$ is independent of RS \cite{RG}, but both the couplant 
$a$ and the coefficients $r_i$ depend on the arbitrary choice of RS\@.  
In particular, $a$ depends on the arbitrary renormalization scale $\mu$:  
\BE
\mu \frac{da}{d\mu} = \beta(a) = - ba^2 B(a),
\EE
where 
\BE
B(a)=1+ca + c_2a^2+c_3a^3+\ldots.
\EE
The first two coefficients of the $\beta$ function are RS invariant and are given by 
\BE
b = \frac{(33-2n_f)}{6}, \quad\quad c= \frac{153 - 19 n_f}{2(33-2n_f)}.
\EE
The higher $\beta$-function coefficients $c_2, c_3, \ldots$ are RS dependent:   
they, together with $\mu/{\tilde{\Lambda}}$, can be used to parametrize the 
RS choice \cite{OPT}.  Certain combinations of ${\cal R}$ and $\beta$-function coefficients are 
RS invariants \cite{OPT}.  (Their definition, and that of $\tilde{\Lambda}$, will 
be discussed in Sec.~\ref{OPTsect}.)  The first few are:
\BA
\tilde{\rho}_1 & = & c, \quad \quad {\mbox{\rm and}} \quad\quad 
\boldsymbol{\rho}_1(Q)= b \ln(\mu/{\tilde{\Lambda}})-r_1 ,
\nonumber \\
\tilde{\rho}_2 & = & c_2+r_2-c r_1-r_1^2, 
\label{rhodefs}
\\
\tilde{\rho}_3 & = & c_3 + 2 r_3 -2 c_2 r_1 -6 r_2 r_1 + c r_1^2 + 4 r_1^3 .
\nonumber 
\EA
The $\tilde{\rho}_i$ are $Q$ independent, since the $\mu/Q$ dependence from the 
$r_i$'s cancels out.   The special invariant $\boldsymbol{\rho}_1(Q)$ depends on 
$Q$ and can be written as 
\BE 
\boldsymbol{\rho}_1(Q) = b \ln(Q/ \tilde{\Lambda}_{{\cal R}}),
\EE
where $\tilde{\Lambda}_{{\cal R}}$ is a scale specific to the particular physical 
quantity ${\cal R}$.

\section{BZ expansion in `regular' schemes}
\setcounter{equation}{0}
\label{BZsect}
                                                                                                                                                                                                                                                                                                                                                                                                                                                                                                                                        
At $n_f=\frac{33}{2}=16\half$ the leading $\beta$-function coefficient $b$ vanishes.                                                                                                                                                                                                                                                                                                                                                                                                                                                                                                                                                For $n_{f}$ just below $16 \half$ the $\beta$ function has a zero at a very small $a^*$, 
proportional to $(16 \half - n_{f})$.  Its limiting form,
\BE
a_{0} \equiv \frac{8}{107} b =\frac{8}{321}\left( 16 \half - n_{f}\right) ,
\EE
serves as the expansion parameter for the Banks-Zaks (BZ) expansion \cite{BZ}-\cite{BZlett}.  
To proceed, one first re-writes all perturbative coefficients, 
eliminating $n_f$ in favour of $a_0$.  The first two $\beta$-function 
coefficients, which are RS invariant, become:
\begin{eqnarray}
b &=& \frac{107}{8} a_{0},  \\
c &=& -\frac{1}{a_{0}} + \frac{19}{4}. 
\end{eqnarray}
Note that $c$ is large and {\it negative} in the BZ context.  

     We will consider a class of physical quantities (dubbed `primary' quantities) for 
which the $\tilde{\rho}_i$ invariants have the form
\BE
\label{rhoiexp}
\tilde{\rho}_{i} = \frac{1}{a_{0}} \left( \rho_{i,-1} + \rho_{i,0} a_{0} 
+ \rho_{i,1} a_{0}^2 + \ldots \right).  
\EE    
Within the class of so-called `regular' schemes \cite{grun,BZlett}, the $\beta$-function 
coefficients  $(b c_i)$ are analytic in $a_0$ so that 
\BE
\label{ciexp}
c_i = \frac{1}{a_{0}} \left( c_{i,-1} + c_{i,0} a_{0} + c_{i,1} a_{0}^2 
+ \ldots \right) .
\EE
Note that this equation is a property of the scheme, irrespective of the physical quantity, 
whereas Eq.~(\ref{rhoiexp}) is a property of the physical quantity, irrespective of 
the scheme.  For `primary' quantities in `regular' schemes we have 
\BE 
\label{riexp}
r_i = r_{i,0} + r_{i,1} a_0 + r_{i,2} a_0^2 + \ldots .
\EE

[In fact, for certain quantities the numerator of Eq.~(\ref{rhoiexp}) is a polynomial whose 
highest term is $\rho_{i,i} {a_0}^{i+1}$, and in certain `rigid' schemes, such as $\msbar$, 
the $a_0$ series for $c_i$ and $r_i$ truncate after the $c_{i,i-1}$ and $r_{i,i}$ terms.  These 
properties are unimportant here, but are crucial in the opposite limit, the large-$b$ 
approximation.]

     Expanding in powers of $a_0$ the zero of the $\beta$ function is found to be
\BE
     a^*=a_0 \left( 1+ (c_{2,-1} + c_{1,0}) a_0 + \ldots \right),
\EE
and hence the infrared limit of ${\cal R}$ is 
\BE
{\cal R}^* = a_0 \left( 1+ (r_{1,0}+c_{2,-1}+c_{1,0}) a_0 + \ldots \right).
\EE
Since the BZ expansion parameter $a_0$ is RS invariant the coefficients in the 
${\cal R}^*$ series are RS invariant and can be written in terms of the 
$\rho_{i,j}$:
\BE
{\cal R}^* = a_0 \left( 1+ (\rho_{2,-1}+\rho_{1,0}) a_0 + \ldots \right).
\EE
Note, though, that $a^*$ is not a physical quantity and its $a_0$ expansion has 
RS-dependent coefficients.

     At a finite energy $Q$ the result for ${\cal R}$ to $n$th order of the BZ expansion
can be expressed as the solution an equation of the form \cite{BZlett}
\BE 
\label{eq: YY}
\boldsymbol{\rho}_1(Q) = \frac{1}{\cal R} + \frac{1}{\hat{\gamma}^{*(n)}} 
\ln\left(1- \frac{{\cal R}}{{\cal R}^{*(n)}} \right) 
+ c\ln\left(\left|c\right| {\cal R} \right) 
\EE
for $n=1,2,3$.  (For $n \ge 4$ there are additional terms; see Ref.~\cite{BZlett} for details.) 
Here ${\cal R}^{*(n)}$ and $\hat{\gamma}^{*(n)}$ are the $n$th-order approximations 
to ${\cal R}^*$ and $\hat{\gamma}^*\equiv \frac{\gamma^*}{b}$.  The critical 
exponent $\gamma^*$ governs the manner in which ${\cal R}$ approaches ${\cal R}^*$ 
in the $Q \to 0$ limit: 
\BE
({\cal R}^*-{\cal R}) \propto Q^{\gamma^*}.
\EE  
Normally $\gamma^*$ is identified with the slope of the $\beta$ function 
at the fixed point \cite{Gross}, and that is true in the present context.  
(Some subtleties with $\gamma^*$ \cite{chylafp,effexp} are discussed in Appendix A.)  
The BZ expansion of $\gamma^*$ is 
\BE
\hat{\gamma}^* \equiv \frac{\gamma^*}{b} = a_0 \left( 1+ g_1 a_0 
+ g_2 a_0^2 + O(a_0^3) \right),
\EE
where the $g_i$'s are the universal invariants of Grunberg \cite{grun}:
\BA
g_1=c_{1,0} &=&\rho_{1,0}, \nonumber \\
g_2=c_{1,0}^2 -c_{2,-1}^2-c_{3,-1} &=& \rho_{1,0}^2 -\rho_{2,-1}^2-\rho_{3,-1} .
\EA
They are universal in that they do not depend on the specific physical quantity ${\cal R}$ 
being considered, and invariant because they can be expressed as combinations of the 
invariants $\rho_{i,j}$ (combinations in which all the $r_{i,j}$ terms cancel).

    Close to the BZ limit ${\cal R}$ remains almost constant over a huge range of $Q$ about 
$\tilde{\Lambda}_{{\cal R}}$.  This constant value is not ${\cal R}^*$ but $0.78 {\cal R}^*$ 
\cite{BZlett}.   More precisely, it is ${\cal R}^*/(1+\chi)$ where $\ln \chi + \chi +1=0$, a 
result that follows from Eq.~(\ref{eq: YY}) to leading order in $a_0$ with 
$\boldsymbol{\rho}_1(Q)=0$, corresponding to $Q=\tilde{\Lambda}_{{\cal R}}$.   
Only when $Q/\tilde{\Lambda}_{{\cal R}}$ becomes extremely small does 
${\cal R}$ abruptly rise up to ${\cal R}^*$, and only when $Q/\tilde{\Lambda}_{{\cal R}}$ 
becomes extremely large does ${\cal R}$ very slowly decrease to zero, as required by 
asymptotic freedom.   (See Fig.~1.)

\begin{figure}[!hbt]

\centering
\includegraphics[width=0.59 \textwidth]{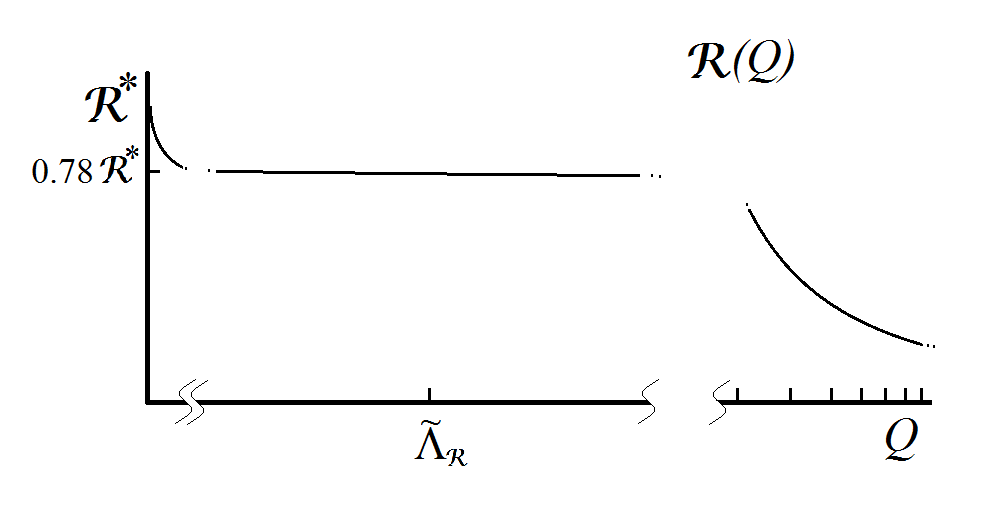}
\vspace*{-2mm}
\begin{quote}

{\setlength{\baselineskip}{0.8\baselineskip} 
Fig. 1.  Schematic picture of ${\cal R}$ as a function of $Q$ close to the BZ limit showing 
the three regions (i) the ``spike'' at very low energies, (ii) the huge flat region where the theory 
is ``nearly scale invariant,'' and (iii) the slow approach to asymptotic freedom at very high energies. 
(Region (iii) is shown on a log scale.) 
 \par}
 
\end{quote}
\end{figure}

   Since Eq.~(\ref{eq: YY}) completely characterizes the $Q$ dependence of ${\cal R}$ in 
low-orders of the BZ expansion, it suffices to consider ${\cal R}^*$ and $\hat{\gamma}^*$, 
both of which are quantities defined in the $Q \to 0$ limit.

\section{Optimized perturbation theory}
\setcounter{equation}{0}
\label{OPTsect}

Since it is a physical quantity, ${\cal R}$ satisfies a set of RG 
equations~\cite{OPT}
\BA
\label{rga}
 \frac{\partial {\cal R}}{\partial \tau} =
\left( \left. \frac{\partial}{\partial \tau} \right|_a + 
\frac{\beta(a)}{b} \frac{\partial}{\partial a} \right) {\cal R} \, = 0, 
 & \quad\quad ``j=1{\mbox{\rm''}}, & \nonumber \\ 
 & & \\
\frac{\partial {\cal R}}{\partial c_j} =
\left( \left. \frac{\partial}{\partial c_j} \right|_a + 
\beta_j(a) \frac{\partial}{\partial a} \right) {\cal R} = 0, & \quad\quad\quad 
j = 2,3,\ldots. & \nonumber
\EA
The first of these, with $\tau \equiv b \ln(\mu/\tilde{\Lambda})$, is the familiar 
RG equation expressing the invariance of ${\cal R}$ under changes of renormalization 
scale $\mu$.  The other equations express the invariance of ${\cal R}$ under other 
changes in the choice of RS\@.  The $\beta_j(a)$ functions, defined as 
$\partial a/\partial c_j$, are given by \cite{OPT,OPTnew} 
\BE
\beta_j(a) \equiv \frac{a^{j+1}}{(j-1)} B_j(a),
\EE
with
\BE
\label{Bj}
B_j(a) = \frac{(j-1)}{a^{j-1}} B(a) I_j(a),
\EE
where 
\BE
\label{Ij}
I_j(a) \equiv \int_0^a dx \, \frac{x^{j-2}}{B(x)^2}.
\EE
The $B_j(a)$ functions have expansions that start $1+O(a)$.  (Note that 
for $j \to 1_+$ one naturally finds $B_1(a)=B(a)$.)

     As mentioned earlier, certain combinations of $r_i$ and $c_j$ coefficients 
form the RS invariants $\tilde{\rho}_i$.  (See Eq.~(\ref{rhodefs})).   
Dependence on $Q$ enters only 
through $\boldsymbol{\rho}_1(Q)=b \ln (Q/\tilde{\Lambda}_{{\cal R}})$.  
The scale $\tilde{\Lambda}_{\cal R}$ is related by 
$\tilde{\Lambda}_{\cal R}=\tilde{\Lambda} \exp ({r_1{\scriptstyle (\mu=Q)}/b)}$ 
to a universal but RS-dependent $\tilde{\Lambda}$ parameter that arises as the 
constant of integration in the integrated $\beta$-function equation:
\BE
\label{intbeta}
 b \ln(\mu/\tilde{\Lambda}) \equiv \tau = K(a),
\EE
where 
\BE
\label{K}
K(a) \equiv \frac{1}{a} + c \ln (\mid\!c\!\mid\! a) - 
\int_0^a \frac{dx}{x^2} \left( \frac{1}{B(x)}-1 + c x \right).
\EE
(This form of $K(a)$, completely equivalent to our previous definition 
\cite{OPT,OPTnew}, is more convenient when $c$ is negative \cite{unfixed}.)    
The $\tilde{\Lambda}$ parameter thus defined is RS dependent, but it 
can be converted between different schemes {\it exactly} by the 
Celmaster-Gonsalves relation~\cite{CG}.

The $\beta$ function is RS dependent.  The conversion between two schemes 
(primed and unprimed) is given by
\BE
\beta'(a') \equiv \mu \frac{da'}{d\mu} = \frac{da'}{da} \mu \frac{da}{d\mu} = \frac{da'}{da} \beta(a).
\EE
For any specific physical quantity ${\cal R}$ one can always define the 
``fastest apparent convergence''  (FAC) or ``effective charge'' (EC) scheme \cite{Grunberg} in 
which all the series coefficients $r_i$ vanish, so 
that ${\cal R} = a_{\EC} (1+0+0+\ldots)$.  As a special case of the previous equation we have
\BE 
\beta_{\EC}({\cal R}) = \frac{d{\cal R}}{da} \beta(a).
\EE
The $\tilde{\rho}_n$ invariants can conveniently be defined to coincide with the 
coefficients of the EC $\beta$ function.   Thus, defining 
$\beta_{\EC}({\cal R}) = - b {\cal R}^2 B_{\EC}({\cal R})$, with
\BE
B_{\EC}({\cal R}) \equiv \sum_{n=0}^{\infty} \tilde{\rho}_n {\cal R}^n,
\EE
the invariants $\tilde{\rho}_n$ can be obtained by equating coefficients in 
\BE 
\label{invariantsmastereq}
B_{\EC}({\cal R}) = \frac{a^2}{{\cal R}^2} \frac{d{\cal R}}{da} B(a),
\EE
which we shall refer to as the ``invariants master equation.''  

   The $\kplus$-order approximation, ${\cal R}^{(k+1)}$, in some general RS, 
is defined by truncating the ${\cal R}$ and $\beta$ series after the $r_k$ and 
$c_k$ terms, respectively:
\BE
{\cal R}^{(k+1)} \equiv a \sum_{m=0}^{k} r_m a^m, \quad\quad 
B^{(k+1)} \equiv \sum_{j=0}^{k} c_j a^j ,
\EE
with $r_0 \equiv 1$, $c_0 \equiv 1$, and $c_1 \equiv c$.  
  Because of these truncations, the resulting approximant depends on RS\@.  
``Optimization'' \cite{OPT} corresponds to finding the stationary point where 
the approximant is locally insensitive to small RS changes, i.e., finding the 
``optimal'' RS in which the RG equations (\ref{rga}) are satisfied by 
${\cal R}^{(k+1)}$ with no remainder.  The resulting optimization equations 
\cite{OPT} have been solved for the optimized $\bar{r}_m$ coefficients 
in terms of the optimized couplant $\bar{a}$ and the optimized $\bar{c}_j$ 
coefficients \cite{OPTnew}.  (The overbars denote quantities in the optimal scheme, 
but we will generally omit these henceforth, except to distinguish $\bar{a}$ from a 
generic $a$.)  To present that solution it is convenient to define
\BE
{\cal S} \equiv \frac{d{\cal R}}{da}=1+s_1 a+ s_2 a^2 + \ldots,
\EE
with coefficients $s_m \equiv (m+1)r_m$.  The optimized $s_m$ 
coefficients are given by \cite{OPTnew}:  
\BE 
\label{formula}
s_m \bar{a}^m = 
\frac{1}{B_k(\bar{a})} \left( H_{k-m}(\bar{a}) - H_{k-m+1}(\bar{a}) \right), 
\quad\quad\quad m=0,1,\ldots,k,
\EE 
where 
\BE
\label{Hdef}
H_i(a) \equiv \sum_{j=0}^{k-i} c_j a^j \left( \frac{i-j-1}{i+j-1} \right) 
B_{i+j}(a), 
\quad\quad\quad i=(1),2,\ldots,k,
\EE
with $c_0 \equiv 1$, $c_1\equiv c$.  $H_1$ is to be understood as the limit 
$i \to 1$ of the above formula.  Note that $H_k=B_k$ and we define 
$H_0 \equiv 1$ and $H_{k+1} \equiv 0$.  

    As noted at the end of the last section, we may focus on the infrared limit $Q \to 0$.  
A finite infrared limit in optimized perturbation theory (OPT) can occur in at least 
two ways: (i) through a fixed point (a zero of the optimized $\beta$ function) 
\cite{KSS,CKL,lowen} or (ii) through an ``unfixed point''  and the pinch mechanism 
\cite{unfixed}.  The latter case is discussed in Appendix B, but seems to be only 
tangentially relevant in the BZ limit.  

    In the fixed-point case the infrared limit of the optimized couplant is $a^*$, which is 
the first zero of the optimized $\beta$ function: $B^{(k+1)}(a^*)=0$.  The above solution for the 
optimized $s_m\equiv (m+1) r_m$ coefficients in terms of the optimized $c_j$ coefficients 
simplifies greatly to 
\cite{OPTnew}
\BE
\label{fpOPTcond}
\hat{s}_m = \frac{1}{(k-1)} \left[ (k-2m) \hat{c}_m -  \sum_{j=0}^{m} \hat{c}_j  \right],
\EE
where $\hat{s}_m \equiv s_m {a^*}^m$, and $\hat{c}_m \equiv c_m {a^*}^m$.

                                                                                                                                                                                                                                                                                                                                                                                                                                                                                                                                \section{Low orders of OPT in the BZ limit}
\setcounter{equation}{0}
\label{Loword}

    Explicit results for the infrared-fixed-point limit of OPT were obtained in Ref. \cite{KSS} 
for $k=2$ and $3$.  Extending the calculation to higher orders is made easier by the 
formula (\ref{fpOPTcond}), which can be used to substitute for the optimal-scheme $r_m$'s 
in the $\tilde{\rho}$ invariants.  From the resulting $\tilde{\rho}_2$ expression one can solve 
for the optimal-scheme $c_2$ in terms of $a^*, c, \tilde{\rho}_2$.   Then, making use of that 
result, one may solve for $c_3$ in terms of $a^*, c, \tilde{\rho}_2, \tilde{\rho}_3$, and so on 
up to $c_{k-1}$.  The last coefficient, $c_k$ can then be found from the fixed-point condition 
$B(a^*)=0$. Substituting in the expression for $\tilde{\rho}_k$ then produces an equation 
for $a^*$ that involves only the invariants $c, \tilde{\rho_2}, \ldots, \tilde{\rho}_k$.   One can 
then find $a^*$ numerically as the smallest positive root of that equation.  Finally, the expressions 
for the $c_j$'s in terms of $a^*$ and the invariants can be substituted in the formula 
(\ref{fpOPTcond}) to determine the $r_m$'s.  Hence, one can find ${\cal R}^*$.  

     The preceding discussion pre-supposes that the perturbative calculations have been done 
to $\kplus$ order, so that the numerical values of the invariants up to $\tilde{\rho}_k$ are 
known.  The great simplification in the BZ limit is that we can effectively set almost all 
the invariants to zero: this can be seen as follows.  As $a_0 \to 0$ the most singular term 
in any of the $\tilde{\rho}_i$ is of order $1/a_0$, but each $\tilde{\rho}_i$ enters the analysis 
along with a factor of ${a^*}^i$ that is of order $a_0^i$.   Thus, to find the leading term 
in the BZ limit, we can effectively set to zero all the invariants except $c$.  (Furthermore, only 
the $-1/a_0$ piece of $c$ will contribute.)  To obtain the next-to-leading correction 
in $a_0$ we would also need the $\frac{19}{4}$ piece of $c$ along with the $\rho_{2,-1}/a_0$ 
piece of $\tilde{\rho}_2$ (whose value depends on the specific ${\cal R}$ quantity 
under consideration).

    For $k=2$, following the procedure in the first paragraph of this section, we find
\BE
r_1=-\frac{1}{2} \left( \frac{1+ c a^*}{a^*} \right), \quad r_2=-\frac{2}{3} c_2,
\EE
from the optimization condition, Eq.~(\ref{fpOPTcond}).  Then $c_2$ can be found from
$B(a^*)=0$ as 
\BE 
c_2 = - \frac{1+c a^*}{{a^*}^2}.
\EE
Substituting in the expression for $\tilde{\rho}_2$ in Eq.~(\ref{rhodefs}) yields the 
equation for $a^*$:
\BE
-\frac{7 + 4 c a^* - 3 c^2 {a^*}^2}{12 {a^*}^2} = \tilde{\rho}_2 .
\EE
(When comparing with Refs. \cite{KSS,lowen} note that the ``$\rho_2$'' used there 
is $\tilde{\rho}_2-\quarter c^2$).  In the BZ limit we can set $\tilde{\rho}_2=0$ so that 
the $a^*$ equation becomes
\BE
(c a^*+1)(c a^* - {\textstyle \frac{7}{3}})=0.
\EE
Hence, we find $a^*=-1/c \to a_0$.  The coefficients $c_2$, $r_1$, $r_2$ all vanish,  
so, in an {\it a posteriori} sense, the $k=2$ OPT scheme is `regular' in the infrared 
(fixed-point) limit.  The final result for ${\cal R}^*$ is
\BE
{\cal R}^* = - \frac{1}{c} \,\,\, \to a_0.
\EE
Thus, exactly as in any `regular' scheme, we find that $a^*$ and ${\cal R}^*$ tend 
to $a_0$ in the BZ limit.  The same is true for $\hat{\gamma}^*$, obtained from the 
slope of the $\beta$ function at the fixed point.

      At higher orders, though, the OPT scheme is not `regular'  --- the optimized 
$r_i$ coefficients, for instance, have $1/a_0^i$ pieces --- and the story is more 
complicated.  For $k=3$ the optimization condition gives
\BE
r_1= - \frac{1}{4 a^*},  \quad     r_2 =-\frac{(1 + c a^* +2 c_2 {a^*}^2)}{6 {a^*}^2}, 
\quad 
r_3 = - \frac{3}{8} c_3.
\EE
Proceeding immediately to the BZ limit, we set $\tilde{\rho}_2=\tilde{\rho}_3=0$.  
Substituting into $\tilde{\rho}_2=0$ gives
\BE
c_2=\frac{(11 - 4 c a^*)}{32 {a^*}^2},
\EE
and then the last coefficient, $c_3$ can be found from $B(a^*)=0$; after using the 
previous equation, this gives
\BE
c_3 = - \frac{(43 + 28 c a^*)}{32 {a^*}^3}
\EE
The equation for $a^*$ in the BZ limit then follows by substituting in $\tilde{\rho}_3=0$.  
We could have expected a cubic equation, but in fact we find
\BE
83 + 52 c a^* =0.
\EE
Thus, we do {\it not} get $a^* =-\frac{1}{c} \to a_0$, but 
$a^* \to \frac{83}{52} a_0 = 1.596 a_0$.   The final result for 
${\cal R}^*$ is not $a_0$ but is $\frac{6889}{6656} a_0 = 1.035 a_0$, which is 
remarkably close.  

   Results for higher orders are shown in Tables~\ref{OPTEtable} and \ref{OPTOtable}.   
The even-$k$ results are significantly better than those for odd $k$.  Note that 
$a^*/a_0$ increases, apparently towards $4$.  It is perfectly acceptable for 
$a^*$ to differ from $a_0$, since $a^*$ is inherently scheme dependent.  However, 
${\cal R}^*$ is a physical quantity so it is reassuring that ${\cal R}^*/a_0$ is always 
close to $1$.  In Sect.~\ref{BZNLS} we will find a simple explanation for 
$a^*/a_0 \to 4$ and ${\cal R}^*/a_0 \to 1$ as $k \to \infty$.

\begin{table}[htbp]
\begin{center}
\begin{tabular}[b]{|r||d{5}|d{5}|d{4}|}
\hline
$k$ \quad & 
\multicolumn{1}{c|}{$\frac{a^*}{a_0}$} & 
\multicolumn{1}{c|}{$\frac{{\cal R}^*}{a_0}$} & 
\multicolumn{1}{c|}{$\frac{\hat{\gamma}^*}{a_0}$} \\
\hline
$2$    &  1            &  1            &  1         \\
$4$    &  1.85035  &  1.00370  &  0.9841 \\
$6$    &  2.30294  &  1.00214  &  0.9742 \\
$8$    &  2.58980  &  1.00137  &  0.9671 \\
$10$  &  2.78928  &  1.00096  &  0.9614 \\
$12$  &  2.93666  &  1.00071  &  0.9565 \\
$14$  &  3.05030  &  1.00055  &  0.9523 \\
$16$  &  3.14081  &  1.00043  &  0.9485 \\
$18$  &  3.21470  &  1.00035  &  0.9451 \\
\hline

\end{tabular}
\end{center}
\vspace*{-5mm}
\caption{ {\textsl{OPT results in the BZ limit for $k=$~even.
}}}

\label{OPTEtable}
\end{table}

\vspace*{3mm}

\begin{table}[htbp]
\begin{center}
\begin{tabular}[b]{|r||d{5}|d{5}|d{4}|}
\hline
$k$ \quad & 
\multicolumn{1}{c|}{$\frac{a^*}{a_0}$} & 
\multicolumn{1}{c|}{$\frac{{\cal R}^*}{a_0}$} & 
\multicolumn{1}{c|}{$\frac{\hat{\gamma}^*}{a_0}$} \\
\hline
$3$    &  1.59615  &  1.03501  &  0.5602  \\
$5$    &  2.17343  &  1.01119  &  0.5886  \\
$7$    &  2.51313  &  1.00544  &  0.6071  \\
$9$    &  2.73950  &  1.00319  &  0.6206  \\
$11$  &  2.90228  &  1.00209  &  0.6311  \\
$13$  &  3.02550  &  1.00147  &  0.6397  \\
$15$  &  3.12231  &  1.00108  &  0.6468  \\
$17$  &  3.20056  &  1.00083  &  0.6530  \\
$19$  &  3.26522  &  1.00066  &  0.6583  \\
\hline

\end{tabular}
\end{center}
\vspace*{-5mm}
\caption{ {\textsl{OPT results in the BZ limit for $k=$~odd.
}}}

\label{OPTOtable}
\end{table}

    The situation with $\hat{\gamma}^*$ is less clear.  This is also a physical quantity 
(with the caveats of Appendix A) so we should have $\hat{\gamma}^*/a_0 \to 1$ as 
$k \to \infty$.  The numerical results in the tables cannot be said to support that contention, 
but neither are they inconsistent with it; one can make good fits to the data with functions 
of $k$ that very slowly approach $1$ as $k=\infty$ for both even and odd $k$.  

    It is hard to go to much larger $k$ with the method described in this section, so we turn to 
an analytic approach in the next sections.  Our results -- albeit in approximations to OPT rather 
than true OPT -- support the claim that $a^*/a_0 \to 4$ and that both ${\cal R}^*/a_0$ and 
$\hat{\gamma}^*/a_0$ tend to $1$ as $k \to \infty$: they also provide valuable insight into 
the workings of OPT at arbitrarily high orders.

                                                                                                                                                                                                                                                                                                                                                                                                                                                                                                                                        \section{Analytic tools for OPT at all orders}                                                                                                                                                                                                                                                                                                                                                                                                                                                                                                                                        \setcounter{equation}{0}                                                                                                                                                                                                                                                                                                                                                                                                                                                                                                                                                                                                                                                                                                                                                                                                                                                                                                                                                                                                                                                                             \label{ExpOPT}
                                                                                                                                                                                                                                                                                                                                                                                                                                                                                                                                                                                                                                                                                                                                                                                                                                                                                                                                                                                                                                                                                                                                                                                                                                                                                                                                                                                                                                                                                                                                                                                                                                                                                                                                                                                                                                                   To make progress analytically with OPT in $\kplus$ order it helps greatly to 
deal with functions and differential equations rather than with $2k$ individual $r_i$ 
and $c_i$ coefficients.  
The set of $\tilde{\rho}_i$ invariants naturally follow from a single ``master equation,'' 
Eq.~(\ref{invariantsmastereq}), and what we need is to also formulate the $k$ optimization 
conditions as a ``master equation.''  For general $Q$ this would be a daunting task.  In the 
infrared fixed-point limit, however, it is relatively simple --- and, happily, that suffices in the 
present context since, as noted in Sect.~\ref{BZsect}, in the BZ limit and for the first three 
terms of the BZ expansion, the entire $Q$ dependence of ${\cal R}$ is characterized by the 
two infrared quantities ${\cal R}^*$ and $\hat{\gamma}^*$.

     We now show that the optimization conditions in the fixed-point limit, Eq.~(\ref{fpOPTcond}), 
follow from equating coefficients in the following ``fixed-point OPT master equation:''
\BE
\label{fpOPTmastereq}
\frac{d{\cal R}}{da}= B(a)- \frac{a}{(k-1)} \left( 2 \frac{d B(a)}{da} + \frac{B(a)}{(a^*-a)} \right) .
\EE
(Superscripts ``${\scriptstyle{(k+1)}}$'' on ${\cal R}$ and $B(a)$ are omitted for brevity.)   
Note that $a$ here is merely a dummy variable, while $a^*$ is the optimized couplant in the 
infrared limit.  

    The first step of the proof is to note that, by the definition of $a^*$, the polynomial $B(a)$ 
has a factor of $a^*-a$ and can be written as 
\BE
\label{Bfactorized}
B(a) = \frac{(a^*-a)}{a^*} \sum_{n=0}^{k-1} \left( \frac{a}{a^*} \right)^n \hat{t}_n,
\EE
where $\hat{t}_n$ is a partial sum of $\beta$-function terms:  
\BE
\hat{t}_n = \sum_{j=0}^{n} \hat{c}_j 
\EE
with $\hat{c}_j \equiv c_j {a^*}^j$.  Note that $\hat{t}_n-\hat{t}_{n-1} = \hat{c}_n$ 
and that $\hat{t}_k=0$ by virtue of the fixed-point condition.  
To show Eq.~(\ref{Bfactorized}), expand the right-hand side, then use $\hat{t}_k=0$ and 
define $\hat{t}_{-1} \equiv 0$ to get 
\BE 
\sum_{n=0}^{k} \left( \frac{a}{a^*} \right)^n \hat{t}_n  - 
\sum_{n=-1}^{k-1} \left( \frac{a}{a^*} \right)^{n+1} \hat{t}_n .
\EE
Now put $n=n'-1$ in the second sum and recombine the sums to get 
\BE
\sum_{n=0}^{k} \left( \hat{t}_n-\hat{t}_{n-1} \right) \left( \frac{a}{a^*} \right)^n 
= \sum_{n=0}^{k} \hat{c}_n \left( \frac{a}{a^*} \right)^n 
= \sum_{n=0}^{k} c_n a^n,
\EE
which is $B(a)$, as claimed.  

     To prove Eq.~(\ref{fpOPTmastereq}), equate powers of $(a/a^*)^m$, using 
Eq.~(\ref{Bfactorized}) to write $B(a)/(a^*-a)$ as a polynomial.  This leads to
\BE 
\hat{s}_m =\hat{c}_m - \frac{1}{(k-1)} \left( 2m \hat{c}_m + \hat{t}_{m-1} \right). 
\EE
Using $\hat{t}_m-\hat{t}_{m-1}=\hat{c}_m$ again and simplifying leads to the fixed-point 
optimization conditions, Eq.~(\ref{fpOPTcond}), completing the proof. 

     Unfortunately, Eq.~(\ref{fpOPTmastereq}) proves difficult to deal with.  
To make progress we have resorted to two approximations, designated PWMR and NLS, 
that we now explain.  Ref.~\cite{OPTnew} has shown that the series expansion of 
$H_i(a)-1$ starts 
\BE 
 H_i(a) - 1 = \frac{k-2i+2}{k} c_{k-i+1} a^{k-i+1} \left( 1+O(a) \right) ,
\EE
which quickly leads to
\BE
\label{PWMR}
s_m = \frac{k-2m}{k} c_m + O(\bar{a}),
\EE
a result first obtained (in a quite different manner) by 
Pennington, Wrigley, and Minaco and Roditi (PWMR) \cite{PWMR}.  
Dropping the $O(\bar{a})$ term leads to the PWMR approximation 
which is easily formulated as a ``master equation'':
\BE
\label{PWMRmastereq}
\frac{d{\cal R}}{da}= B(a)- \frac{2}{k} a \frac{d B(a)}{da} 
 \quad\quad\quad {\mbox{\rm (PWMR)}}.
\EE
          
    Looking at the above equation, or the original equation (\ref{fpOPTmastereq}), it is 
tempting to suppose that, as $k \to \infty$, they reduce to 
\BE 
\label{NLSmastereq}
\frac{d{\cal R}}{da}= B(a)  \quad\quad\quad {\mbox{\rm (NLS)}}.
\EE
We shall refer to this as the ``na\"{\i}ve limiting scheme" (NLS)\@.  It corresponds 
to a well-defined RS in which $s_m=c_m$, so that the coefficients $r_m=s_m/(m+1)$ of the 
${\cal R}$ series decrease by a factor $1/(m+1)$ relative to the coefficients of the $B$ 
series.                                    

    Clearly, this idea is very na\"{\i}ve.  In the PWMR case the actual relation is 
$s_m=\frac{k-2m}{k} c_m$, which only reduces to $s_m \approx c_m$ for 
$m \ll k$; that is, for the early part of the series only.  Nevertheless, there may be 
a kernel of truth here, for if the series are ``well behaved" the early terms should 
dominate.  In any case, adopting this na\"{\i}ve idea leads us in a fruitful direction.  
Our investigations below will lead us to conclude that, at least in the BZ context, the NLS 
does yield the all-orders limit of OPT, although it is a poor guide to how fast results 
converge to that limit.  
         
     Using the NLS equation above to eliminate $B(a)$ in the invariants master equation 
(\ref{invariantsmastereq}) leads directly to 
\BE 
B_{\EC}({\cal R}) = \frac{a^2}{{\cal R}^2} \left( \frac{d{\cal R}}{da} \right)^2 .
\EE     
Taking the square root leads to 
\BE
\label{NLSkeyEq}
\frac{d{\cal R}}{da}= \frac{{\cal R}}{a} \sqrt{B_{\EC}({\cal R})},
\EE
which is immediately integrable.

       The BZ limit provides us with a nice ``playground" for exploring further, since 
it effectively corresponds to the case  $B_{\EC}({\cal R}) = 1+c {\cal R}$.  
We continue this analysis in the next section.

\section{All-orders NLS in the BZ limit}
\setcounter{equation}{0}
\label{BZNLS}

   In the BZ limit the only one of the $\tilde{\rho}_n$ invariants that contributes 
is $c$, which is negative: $c=-1/a_0 +O(1)$ as $a_0 \to 0$.  We may set 
$B_{\EC}({\cal R})=1+ c {\cal R}$ in this limit.  (The terms neglected can 
only contribute to $O(a_0)$ corrections, as argued in Sect.~\ref{Loword}.)  
It is convenient to define
\BE
\label{udef}
u \equiv \frac{-c a}{4}, \quad\quad\quad v \equiv -c {\cal R}.
\EE
In these variables, the NLS condition is $B= \frac{1}{4} \frac{dv}{du}$ and 
Eq. (\ref{NLSkeyEq}) becomes 
\BE
\frac{dv}{du} = \frac{v}{u} \sqrt{1-v},
\EE
which leads to 
\BE 
\label{NLSvu}
\int \frac{dv}{v \sqrt{1-v}} = \int \frac{du}{u}.
\EE
Performing the integral and then exponentiating both sides gives
\BE 
\frac{1-\sqrt{1-v}}{1+\sqrt{1-v}} =u,
\EE
where the constant of integration has been fixed by requiring 
$v \to 4u$ as $u \to 0$, corresponding to the ${\cal R}$ series 
beginning ${\cal R} = a(1+\ldots)$.  Inverting this equation 
(assuming $u \le 1$) 
gives
\BE
\label{vofu}
v = \frac{4 u}{(1+u)^2}.
\EE
Hence, $B=\frac{1}{4} \frac{dv}{du}$ is given by 
\BE
\label{Bofu}
B= \frac{1-u}{(1+u)^3}.
\EE
(The two formulas above are key results.  They show an interesting $u \to 1/u$ duality 
that we will discuss in Sect.~\ref{dualitySect}.) 

   The fixed point, where $B=0$, is at $u^*=1$.  Recalling Eq.~(\ref{udef}), we see that 
$a^*$ is $-4/c \to 4 a_0$.  Nevertheless, because $u^*=1$ in Eq.~(\ref{vofu}) 
leads to $v^*=1$, we find ${\cal R}^*=-1/c \to a_0$, in agreement with the regular-scheme 
result.   

     Evaluating the slope of the $\beta$ function at the fixed point gives
\BE
-b \left( -\frac{4}{c} \right) u^2 \frac{d}{du} \left. \left( \frac{1-u}{(1+u)^3} \right) \right|_{u=1} = 
\frac{-b}{2 c} \,\,\, \to \frac{b a_0}{2},
\EE
which seemingly gives $\hat{\gamma}^* \equiv \gamma^*/b = \frac{1}{2} a_0$.  
Here the subtlety discussed in Appendix A comes into play.  The critical exponent 
$\gamma^*$ is really the infrared limit of an effective power-law exponent given at 
finite $Q$ by \cite{effexp}
\BE
\label{gamQ7}
\gamma(Q) =   \frac{d \beta}{da} + 
\beta(a) \frac{d^2 {\cal R}}{d a^2} \Big/ \frac{d{\cal R}}{d a} .
\EE
Normally the second term drops out in the infrared limit because $\beta(a)$ 
vanishes at the fixed point.  However, in the NLS the denominator 
$\frac{d {\cal R}}{d a}$ also vanishes because it is $B(a)=\beta(a)/(-b a^2)$.  
Therefore, in the NLS case the second term contributes 
$- b a^2 \frac{d^2 {\cal R}}{d a^2} = -b a^2 \frac{dB}{da}$ which contributes 
an equally with the first term, thus rescaling the previous result by a factor of $2$.  Hence, 
we find $\hat{\gamma}^*=a_0$, in accord with the regular-scheme result.  

      The preceding discussion corresponds to the NLS result re-summed to infinite order.  
One must now ask: Do the finite-order NLS results converge to their infinite-order form 
-- and,  if so, how fast?  At $\kplus$ order the $B$ and $v$ series are truncated, and 
$v^*$ is found by evaluating at $u^*$, the zero of the truncated $B$.   Luckily, as with 
a simple geometric series, the sum of finite number of terms can be expressed fairly 
simply.  The truncated $B$ series is 
\BE
\label{NLSBtrunc}
B^{(k+1)}  =
\sum_{j=0}^k (j+1)^2 (-u)^j =  \frac{1-u}{(1+u)^3} 
+ (-1)^k k^2 \frac{u^{k+1}}{(1+u)} \left( 1+ O(\frac{1}{k}) \right).
\EE
Only for odd $k$ do we get a zero.  (We will discuss even $k$ near the end of this Section.)  
The zero of the truncated $B$ is just before $u$ reaches 1.  If we put 
\BE
u=u^* \equiv 1-\frac{\eta(k)}{k} 
\EE
with $\eta(k) \ll k$, we find (noting that $u^{k+1} \to e^{-\eta(k)}$) that 
\BE
\eta(k) = 3 \ln k - \ln(\ln k) -\ln(3/4) + O\left( \frac{\ln\ln k}{\ln k} \right).
\EE
The truncated $v$ series is
\BE
v^{(k+1)} = 4 u \left( \sum_{j=0}^k (j+1) (-u)^j \right) = 
4 u \left( \frac{1}{(1+u)^2} +(-1)^k k \frac{u^{k+1}}{(1+u)} \left( 1+ O \left( \frac{1}{k} \right) \right) 
\right) .
\EE
When we substitute $u=u^*$ we find a cancellation of the $\eta(k)/k$ terms which leaves 
\BE 
v^* \approx 1- \frac{9}{4} \frac{\ln^2k}{k^2} .
\EE
This is in good accord with the numerical results in Table~\ref{NLStable}.

\begin{table}[htbp]
\begin{center}
\begin{tabular}[b]{|r||d{5}|d{5}|d{2}|}
\hline
$k$ \quad & 
\multicolumn{1}{c|}{$\, 4 u^*=\frac{a^*}{a_0} \,\, $} & 
\multicolumn{1}{c|}{$\, v^*=\frac{{\cal R}^*}{a_0} \,\,$} & 
\multicolumn{1}{c|}{$\frac{\hat{\gamma}^*}{a_0}$} \\
\hline
$3$     &  1.41825  &  0.69455  &  3.67  \\
$11$   &  2.26825  &  0.90345  &  7.14  \\
$19$   &  2.65953  &  0.95010  &  8.79  \\
$51$   &  3.25059  &  0.98737  &  11.70  \\
$101$ &  3.53265  &  0.99555  &  13.66  \\
$601$ &  3.88410  &  0.99976  &  18.71 \\
\hline

\end{tabular}
\end{center}
\vspace*{-5mm}
\caption{ {\textsl{NLS results in the BZ limit.
}}}

\label{NLStable}
\end{table}

A similar analysis for $\hat{\gamma}^*$ (including the factor of $2$ discussed above) 
leads to 
\BE
\hat{\gamma}^* = a_0 \left( 1 + 3 (-1)^{k+1} \ln k + \ldots \right),
\EE
which indicates that the NLS results for $\hat{\gamma}^*$ do {\it not} converge -- 
the nominal limit of $a_0$ is ``corrected'' by a $\ln k$ term arising from the 
series-truncation effects.  We indeed see this in the numerical results in Table~\ref{NLStable}.  

     Returning to Eq.~(\ref{NLSBtrunc}) we see that the truncated $B(u)$ function closely 
approximates its limiting form $\frac{1-u}{(1+u)^3}$ until $u$ gets close to $1$.  For odd 
$k$ the $(-1)^k$ ``truncation effect'' term causes $B$ to suddenly dive down, producing a zero.  
For even $k$ this term causes $B$ to suddenly shoot upwards and there is no zero.  
This means that there is no finite infrared limit in these orders; the ``spike'' in ${\cal R}$ 
goes all the way up to infinity.  However, since $B$ has a minimum very close to zero the 
running of the couplant ``almost stops'' here and if we were to evaluate $v$ at this value of $u$ 
we would find a result close to the ${\cal R}^*/a_0$ obtained in the previous odd-$k$ order.  
A related observation is that, with only a slight change of RS, we would find an infrared limit 
arising from a pinch mechanism (see Appendix B).

   We conclude that the NLS provides a lot of insight into OPT as $k \to \infty$, but is only a 
rather crude approximation to true OPT\@.  We move on to the PWMR approximation in 
the next section.

\section{All-orders PWMR in the BZ limit}
\setcounter{equation}{0}
\label{PWMRSect}

As before we have $B_{\EC}({\cal R})=1+ c {\cal R}$ in the BZ limit and we 
use $u \equiv \frac{-c a}{4}$ and $v \equiv -c {\cal R}$.   In these variables the invariants 
master equation (\ref{invariantsmastereq}) becomes  
\BE
\label{invariantsMeq2}
B= \frac{v^2}{4 u^2} \frac{(1-v)}{\frac{dv}{du}},
\EE
and the PWMR master equation (\ref{PWMRmastereq}) becomes 
\BE
\label{PWMRMeq}
\frac{1}{4} \frac{dv}{du}= B - \frac{2}{k} u \frac{dB}{du}.  
\EE
We will proceed to solve these two coupled differential equations, treating $k$ as an 
ordinary parameter: only later will we consider the other $k$ dependence coming from the 
truncations of the resulting series at $\kplus$ order.   (We have explicitly checked that at 
low $k$ this two-step approach does produce the same results as a PWMR version of the 
OPT procedure described in Sect. \ref{Loword}.)  

     We begin by making an {\it ansatz}:     
\BE
\label{ansatz}
B= \frac{1}{4} \frac{dv}{du} \frac{1}{\xi^2} ,
\EE
where $\xi$ depends on $u$.  (We will actually want to view it as a function of a 
new variable $X$, introduced below, that itself is a function of $u$.)  Substituting in 
Eq.~(\ref{invariantsMeq2}) leads, in the same way as in the NLS case, to 
\BE
\int \frac{dv}{v \sqrt{1-v}} = \int \frac{du}{u} \xi,
\EE
which leads to
\BE
\label{vX}
v = \frac{4 X}{(1+X)^2} ,
\EE
with the new variable $X$ defined by 
\BE 
X \equiv \exp \int \frac{du}{u} \xi,
\EE
or more specifically, enforcing $X \to u$ as $u \to 0$, 
\BE
X \equiv u \exp \int_0^u \frac{d\bar{u}}{\bar{u}} \left( \xi-1 \right).
\EE
Note that 
\BE
\frac{dX}{du} = \frac{X}{u} \xi ,
\EE
so that the inverse relationship is 
\BE
\label{ufromX}
u = X \exp \int_0^X \frac{d\bar{X}}{\bar{X}} \left( \frac{1}{\xi(\bar{X})}-1 \right).
\EE
We will now want to consider $\xi$ as a function of the new variable $X$.  

   We can now find $\frac{dv}{du}$ as $\frac{dv}{dX} \frac{dX}{du}$ and 
substitute back in the {\it ansatz} (\ref{ansatz}) to get 
\BE
\label{Bfromxi}
B = \frac{(1-X)}{(1+X)^3} \frac{X}{u \, \xi} .
\EE
From this we can calculate $\frac{dB}{du}$, which, after some algebra, reduces to 
\BE
\frac{dB}{du}=\frac{B}{u} \left( \frac{(1-4X+X^2)}{(1-X^2)} \xi -1 - X \frac{d\xi}{dX} \right).
\EE  
Substituting this, and $\frac{1}{4} \frac{dv}{du}=\xi^2 B$ from the {\it ansatz} (\ref{ansatz}), 
into Eq.~(\ref{PWMRMeq}), leads, after cancelling a factor of $B$, to an equation for $\xi(X)$:
\BE
\label{xieq}
1-\xi^2 = \frac{2}{k} \left( \frac{(1-4X+X^2)}{(1-X^2)} \xi -1 - X \frac{d\xi}{dX} \right).  
\EE
Remarkably, this nonlinear, first-order differential equation is soluble.  The trick is 
to write $\xi$ in the form
\BE 
\label{xiform}
\xi=1- \frac{2}{k} \frac{X}{{\cal F}} \frac{d {\cal F}}{dX} .
\EE
This substitution, because of a cancellation of $({\cal F}'/{\cal F})^2$ terms, leads to a 
{\it linear} second-order equation for ${\cal F}$.  A further substitution,
\BE
{\cal F} = (1-X)^2 F ,
\EE
leads to a Gauss hypergeometric equation, revealing that 
\BE
F = {}_2F_1(-n,\frac{3}{2},-n-\frac{1}{2}; X^2) ,
\EE
where $n \equiv k/2-1$.  We will focus on the case of even $k$.  (Curiously, the roles of 
odd and even $k$ are reversed relative to the NLS case.)  
For even $k$ the $F$ function is a polynomial of degree $n$ in $X^2$:
\BE
F= \frac{n!}{(2n+1)!!} \sum_{i=0}^n \frac{(2i+1)!!}{i!} \frac{(2(n-i)+1)!!}{(n-i)!} (X^2)^i .
\EE
The first few $F$'s are shown in Table~\ref{Ftable}.  Note the `reflexive' symmetry $i \to n-i$, 
meaning that the coefficients are symmetric about the middle.  In the $n \to \infty$ limit $F$ 
approaches $(1-X^2)^{-3/2}$, except near $X=1$, where its behaviour involves a 
modified Bessel function $I_1$ (see Table~\ref{Ftable}).  

\begin{table}[htbp]
\begin{center}
\begin{tabular}[b]{|c|c||l|}
\hline
$k$ \quad & 
$n$ \quad & 
$\quad F$ \\
\hline
$2$     &  $0$    &  $1$    \\
$4$     &  $1$    &  $1+X^2$  \\
$6$     &  $2$    &  $1+ \frac{6}{5} X^2 + X^4$  \\
$8$     &  $3$    &  $1+\frac{9}{7} X^2 + \frac{9}{7} X^4 + X^6$ \\
$10$   &  $4$    &  $1+ \frac{4}{3} X^2 + \frac{10}{7} X^4 + \frac{4}{3} X^6 + X^8$  \\
\hline
$\infty$  & $\infty$  & \quad $(1-X^2)^{-3/2} \quad\quad\,\,\, (X \neq 1)$ \\
$       $   & $       $  &  \quad  $\sqrt{n^{3}} \, \frac{\sqrt{\pi}}{2} \, 
\frac{e^{-x} I_1(x)}{x} \quad \,\, (X=1-\frac{x}{n})$ \\
\hline

\end{tabular}
\end{center}
\vspace*{-5mm}
\caption{ {\textsl{The first few $F$ polynomials and their form for large $k=2n+2$.
}}}

\label{Ftable}
\end{table}

    To find $u$ in terms of $X$ it is helpful to use another representation of $\xi$, 
namely
\BE
\label{invxiform}
\frac{1}{\xi} = 1- \frac{1}{n+2} \frac{X}{{\cal P}} \frac{d{\cal P}}{dX},
\EE
so that Eq.~(\ref{ufromX}) will immediately lead to
\BE
u= X \, {\cal P}^{-\frac{1}{(n+2)}} .
\EE
Substituting the above form for $\frac{1}{\xi}$ into the $\xi$ equation (\ref{xieq}) leads again to 
a linear equation.  One can verify that this equation is satisfied by setting 
\BE
{\cal P} = (1+X)^4 P
\EE
with 
\BE
P=\frac{1}{(n+1)} \frac{1}{(1+X)} 
\left(  \left[ n+1-(n-1)X \right] F-2(1-X) X^2 \frac{d F}{d(X^2)} \right).
\EE
The numerator turns out to have a $(1+X)$ factor, so that $P$ is 
a polynomial of degree $2n$ in $X$.  The first few $P$'s are shown in table \ref{Ptable}. 
These polynomials also have a `reflexive' property.  

\begin{table}[htbp]
\begin{center}
\begin{tabular}[b]{|c|c||l|}
\hline
$k$ \quad & 
$n$ \quad & 
$\quad P$ \\
\hline
$2$     &  $0$    &  $1$    \\
$4$     &  $1$    &  $1-X+X^2$ \\
$6$     &  $2$    &  $1- \frac{4}{3} X + \frac{26}{15} X^2- \frac{4}{3} X^3 +X^4$  \\
$8$     &  $3$    &  $1-\frac{3}{2} X+\frac{15}{7} X^2-\frac{15}{7} X^3-\frac{15}{7} X^4
-\frac{3}{2} X^5 + X^6$ \\
$10$   &  $4$    &  $1-\frac{8}{5} X+ \frac{12}{5}X^2-\frac{8}{3}X^3+\frac{62}{21}X^4
-\frac{8}{3}X^5+\frac{12}{5}X^6-\frac{8}{5}X^7+ X^8$  \\
\hline
$\infty$  & $\infty$  &  \quad \quad  $(1-X)^{-1/2}(1+X)^{-5/2} \quad\quad\,\, (X \neq 1)$  \\
$       $   & $       $  &  \quad \quad  
$\sqrt{n} \, \frac{\sqrt{\pi}}{4} \, e^{-x} I_0(x) 
\quad \quad\quad\quad\quad\quad (X=1-\frac{x}{n})$ \\
\hline

\end{tabular}
\end{center}
\vspace*{-5mm}
\caption{ {\textsl{The first few $P$ polynomials and their form for large $k=2n+2$.
}}}

\label{Ptable}
\end{table}

     Yet another expression for $\xi$ is 
\BE
\xi= \frac{(1+X)}{(1-X)} \frac{P}{F},
\EE
which can be proved by substituting for $P$ and simplifying to reach Eq.~(\ref{xiform}).  
Using this form of $\xi$ in Eq.~(\ref{Bfromxi}) gives
\BE
B=(1-X)^2 F \, {\cal P}^{-\left( \frac{n+1}{n+2} \right)}.
\EE

     As noted in the tables, both $F$ and $P$ polynomials have simple limits as $k \to \infty$, 
provided that $X \neq 1$.  It is easy to see that $X \to u$ and that all formulas revert to 
their NLS forms in this limit.  Thus, it is clear that $v^*$ must ultimately tend to $1$, so that 
${\cal R}^* = a_0$ in accord with the BZ limit.  

      However, to go further analytically and determine how fast the finite-order PWMR results
approach their infinite-order form is beset with difficulties; the subtleties when $X \sim 1$
are crucial.  The theory of hypergeometric functions when two parameters go to infinity 
\cite{Watson} is formidably complicated.  
Moreover, in any finite order we need to re-express both $B$ and $v$ as series, not in $X$ 
but in $u$; then find $u^*$ from the zero of the truncated $B$ series; and then evaluate 
the truncated $v$ series at $u=u^*$.  Nevertheless, we can explore these issues numerically 
with Mathematica.  We have been able to explore up to $k \approx 100$ and the numerical 
results are presented in Table~\ref{PWMRtable}.  It appears that $v^*$ approaches $1$ 
significantly faster than in the NLS case:   
\BE
v^* \sim 1 - A \frac{\ln k/k_0}{k^2} ,
\EE
with $A \approx 0.08$ and $k_0 \approx 2.5$, roughly.  

\begin{table}[htbp]
\begin{center}
\begin{tabular}[b]{|r||d{5}|d{5}|d{4}|}
\hline
$k$ \quad & 
\multicolumn{1}{c|}{$\, 4 u^*= \frac{a^*}{a_0} \,\,$} & 
\multicolumn{1}{c|}{$\, v^*= \frac{{\cal R}^*}{a_0} \,\,$} & 
\multicolumn{1}{c|}{$\frac{\hat{\gamma}^*}{a_0}$}    \\
\hline
$2$     &  1            &  1            &  1          \\
$4$     &  1.56878  &  0.99743  &  1.0526  \\
$10$   &  2.41100  &  0.99893  &  1.1064  \\
$18$   &  2.88641  &  0.99952  &  1.1371  \\
$50$   &  3.46514  &  0.99990  &  1.1869  \\
$100$ &  3.69257  &  0.99997  &  1.2183 \\
\hline

\end{tabular}
\end{center}
\vspace*{-5mm}
\caption{ {\textsl{PWMR results in the BZ limit.
}}}

\label{PWMRtable}
\end{table}

    The ratio of $v$ to its NLS form $v_{\rm NLS} \equiv \frac{4u}{(1+u)^2}$ stays very 
close to $1$ in the entire relevant range $0<u<u^*$, although it strongly deviates thereafter. 
See Fig.~2.  
\begin{figure}[!hbt]
\centering
\includegraphics[width=0.52 \textwidth]{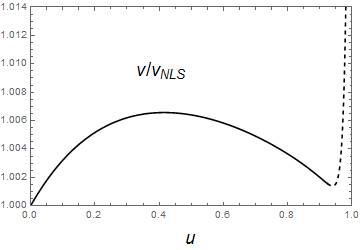}
\vspace*{-2mm}
\begin{quote}

{\setlength{\baselineskip}{0.8\baselineskip} 
Fig. 2.  Plot of $v$ divided by $v_{\rm NLS} \equiv \frac{4 u}{(1 + u)^2}$ as a function 
           of $u$ for PWMR at $k = 100$.  The curve is shown dashed beyond 
           $u = u^* = 0.92314$.
 \par}
 
\end{quote}
\end{figure}

The $v$ series is also much better behaved than in NLS, where the magnitude of the coefficients 
increased in arithmetic progression: $v_{\rm NLS} = 4u \sum_j (j+1) (-u)^j$.  In the PWMR case, 
the coefficients $v_j$ in 
\BE 
\label{vseries}
v = 4u \sum_{j=0}^{k} v_j (-u)^j
\EE
are plotted in Fig.~3 for $k=100$.   
The initial $(j+1)$ growth is suppressed by a more-than-exponential decay (a crude fit is 
$(j+1) \exp (-0.019 j^{3/2})$).  
The middle coefficient $j=\frac{k}{2}$ is exactly zero because of the $k-2j$ factor in the 
PWMR relation between $s_j$ and $c_j$ coefficients, Eq.~(\ref{PWMR}).  The coefficients 
remain very small thereafter.  The somewhat bad behaviour of the last few coefficients is almost 
entirely suppressed by the $u^j$ factor, even at $u=u^*$, the largest relevant $u$, and it actually 
plays a beneficial role.  This can be seen in Fig.~4 which plots the partial sums of 
$n_{\rm max}$ terms of the $v$ series, Eq.~(\ref{vseries}), at $u=u^*$ in the case $k=100$.  
The series has pretty well converged after $50$ terms, but including $25$ more terms significantly 
reduces the error.  The very last term makes an unexpectedly large correction, but this further 
reduces the error and means that the last term provides quite a realistic error estimate.

\begin{figure}[!hbt]
\centering
\includegraphics[width=0.67 \textwidth]{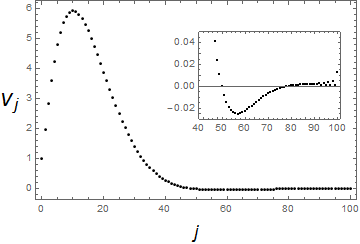}
\vspace*{-2mm}
\begin{quote}

{\setlength{\baselineskip}{0.8\baselineskip} 
Fig. 3.  Coefficients $v_j$ in the series expansion of $v(u)=4u \sum_{j=0}^{k} v_j (-u)^j$, 
           for PWMR with $k=100$.  The inset shows the higher-order coefficients on a finer scale.
 \par}
 
\end{quote}
\end{figure}

%
%

\begin{figure}[!hbt]
\centering
\includegraphics[width=0.59 \textwidth]{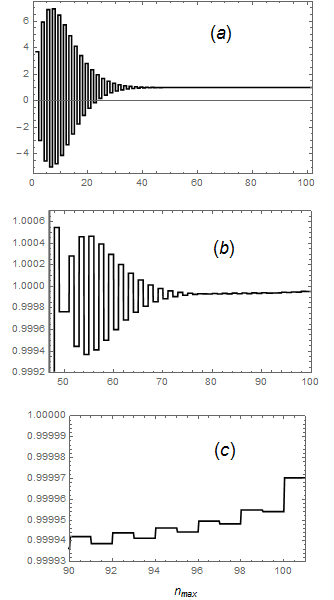}
\vspace*{-2mm}
\begin{quote}

{\setlength{\baselineskip}{0.8\baselineskip} 
Fig. 4.  The partial sums $4 u \sum_ {j = 0}^{n_{\rm max}} v_j (-{u^*})^j$ versus 
           $n_{\rm max}$ for the $v^*$ series in the case $k = 100$.  The plots use three 
           different scales, so as to show that (a) the series has crudely converged after $50$ 
           terms but (b) a slight adjustment from 50 to 75 terms reduces the error quite 
           significantly, and (c) the last term makes an unexpectedly large change, given the 
           trend of the preceding terms, but this further improves the result and means that 
           the last term is, within a factor of $2$, a good measure of the actual error.  
 \par}
 
\end{quote}
\end{figure}

     The series for $\hat{\gamma}^*$, which is just $\left. d\beta/da \right|_*$, is much worse 
behaved.  Also the sequence of results for $\hat{\gamma}^*$ in Table~\ref{PWMRtable} appear to 
diverge, though at a much slower rate than in NLS\@.  It is reasonable to hope that the extra subtleties 
in full OPT would lead to $\hat{\gamma}^*$ converging to $a_0$, albeit very, very slowly, in view of 
the low-orderOPT results in Tables~\ref{OPTEtable} and \ref{OPTOtable}.  

We have not been able to extend the analysis to the full fixed-point master equation, 
(\ref{fpOPTmastereq}).  One can get to an equation similar to Eq.~(\ref{xieq}), but with 
an extra term involving $u/(u-u^*)$ that seems intractable.  Moreover, the parameter $u^*$ 
can only be fixed after the $B(u)$ function is found, and expressed as a truncated series, so 
the interaction between analytic subtleties and truncations effects is even more complicated 
and delicate.

\section{BZ expansion in all-orders OPT}
\setcounter{equation}{0}
\label{BZexpSect}

Setting aside the difficult issue of how fast results converge as $k \to \infty$, the results 
of the last section confirm that the simple NLS formulas from Sect.~\ref{BZNLS}, 
\BE
\label{vofuC}
v = \frac{4 u}{(1+u)^2},
\EE
\BE
\label{BofuC}
B= \frac{1-u}{(1+u)^3},
\EE
represent the all-orders limit of PWMR  -- and presumably of true OPT too -- in the BZ limit.  
As previously noted, these formulas give the same BZ limit for ${\cal R}^*$ and 
$\hat{\gamma}^*$ as `regular' schemes.
We now show that higher terms in the BZ expansion are reproduced 
correctly by all-orders NLS.   

Before discussing the general proof it is instructive to look at next-to-leading order 
in the BZ expansion.  At this level we now need two of the invariants, $c$ and 
$\tilde{\rho}_2$ so we take
\BE
\label{2termBEC}
B_{\EC} = 1+ c {\cal R} + \tilde{\rho}_2 {\cal R}^2 .
\EE
(In fact, only the $\rho_{2,-1}$ piece of $\tilde{\rho}_2$ would contribute 
when we re-expand the results in powers of $a_0$.  However, it will not 
be necessary to carry out that step explicitly, since once we show equivalence to 
the EC scheme, a `regular' scheme, we are bound to get the same BZ expansion to the 
corresponding order in $a_0$.)  Recall that the NLS condition and the invariants 
master equation together lead to Eq.~(\ref{NLSkeyEq}),
\BE
\label{NLSkeyEq2}
\frac{d{\cal R}}{da}= \frac{{\cal R}}{a} \sqrt{B_{\EC}({\cal R})},
\EE
which now gives
\BE
\int \frac{ d{\cal R}}{{\cal R} \sqrt{1+ c {\cal R} + \tilde{\rho}_2 {\cal R}^2 } } 
= \int \frac{da}{a} .
\EE
Integration yields
\BE
 \ln \left( \frac{ {4 \, \cal R}}{2+c {\cal R} + 
 2 \sqrt{1+ c {\cal R} + \tilde{\rho}_2 {\cal R}^2 } } \right) 
 =\ln a,
 \EE
where the constant of integration has been fixed so that ${\cal R} =a (1+\ldots)$ 
as $a \to 0$.  One can now exponentiate and solve for ${\cal R}$, and then $B(a)$ 
can be found from $d{\cal R}/da$.  As before we define $u=-ca/4$ and $v=-c {\cal R}$.  
The zero of $B$ is at 
\BE
u^* = \frac{1}{\sqrt{1- 4 \frac{\tilde{\rho}_2}{c^2}}} ,
\EE
and in terms of these variables we find
\BE
\label{vofu2}
v= \frac{4 u}{\left( 1+ 2 u + \frac{u^2}{{u^*}^2} \right) },
\EE
\BE
\label{Bofu2}
B= \frac{1-\frac{u^2}{{u^*}^2}}{\left( 1+ 2 u + \frac{u^2}{{u^*}^2} \right)^2}.
\EE
It is now straightforward to 
check that $v$ evaluated at $u=u^*$ gives 
\BE
{\cal R}^* = -\frac{v^*}{c} = - \frac{c}{2 \tilde{\rho}_2} 
\left( 1- \sqrt{1- \frac{4 \tilde{\rho}_2}{c^2}} \right),
\EE
which is the root of $B_{\EC}({\cal R}) =0$.   Thus, the ${\cal R}^*$ 
of all-orders NLS agrees with the ${\cal R}^*$ of the EC scheme.  Also, 
$\hat{\gamma}^*$, defined as the infrared limit of Eq.~(\ref{gamQ7}), which leads to 
\BE 
\hat{\gamma}^* = -2 a^2 \left. \frac{dB}{da} \right|_* ,
\EE
with the factor-of-2 subtlety as in Sect.~\ref{BZNLS}, can be shown to reduce to 
\BE
\hat{\gamma}^* = - {\cal R}^2 \left. \frac{dB_{\rm EC}}{d{\cal R}} \right|_* ,
\EE
which is the $\hat{\gamma}^*$ of the EC scheme.  

The general proof is really just a special case of the general formal arguments that 
${\cal R}^*$ and $\hat{\gamma}^*$ (properly defined) are invariant under RS transformations 
\cite{effexp}.  From Eq.~(\ref{NLSkeyEq2}) we can see immediately that $B(a)$, equal to 
$d{\cal R}/da$ in NLS, must vanish when $B_{\EC}$ vanishes; thus the ${\cal R}$ evaluated at 
$a=a^*$ in NLS must agree with the ${\cal R}^*$ defined as the zero of the EC $\beta$ function.  
Furthermore, the equivalence of the two equations for $\hat{\gamma}^*$ above can be proved 
just from the NLS condition $B= d{\cal R}/da$ and Eq.~(\ref{NLSkeyEq2}), without assuming 
any specific form for $B_{\EC}$. 

\section{$\boldsymbol{a} \, {\mathbf{ \to}} \, \boldsymbol{{a^*}^2\!/a}$ duality}
\setcounter{equation}{0}
\label{dualitySect}

It is easily verified that under $u \to {u^*}^2/u$ 
the $v$ of Eq.~(\ref{vofu2}) remains invariant, while the $B$ of Eq.~(\ref{Bofu2}) 
transforms to $-(u^2/{u^*}^2)B$\@.  These properties are even easier to spot in 
Eqs.~(\ref{vofuC}, \ref{BofuC}), in the BZ-limit case, where $u^*=1$.  

Let us try to trace the origin of these properties.  Consider a transformation
\BE
a \longrightarrow \frac{\lambda^2}{a},
\EE
with some positive constant $\lambda$.  We postulate that ${\cal R}$ and all the 
$\tilde{\rho}_i$ invariants remain invariant and that the $\beta$-function equation, 
$\mu \frac{da}{d \mu}=\beta(a)$ maintains its form.  The latter condition means 
that 
\BE
\frac{d a}{d \tau} = -a^2 B(a),
\EE
where $\tau = b \ln(\mu/\tilde{\Lambda})$, must transform to
\BE
\frac{d}{d \tau} \left( \frac{\lambda^2}{a} \right) = - \left( \frac{\lambda^2}{a} \right)^2 
B^{\rm T}(a),
\EE
where $B^{\rm T}(a) \equiv B(\frac{\lambda^2}{a})$.  This requires 
\BE
B^{\rm T}(a) = - \frac{a^2}{\lambda^2} B(a).
\EE
If $B(a)$ vanishes at $a=a^*$ then $B^{\rm T}(a)$ must too.  Thus 
$\lambda^2/a^*$ must be a zero of $B(a)$.  If we assume that there is only one 
zero, then we must take $\lambda=a^*$.  

The transformation of $\frac{d{\cal R}}{da}$ would be
\BE 
\frac{d{\cal R}}{d a} \longrightarrow \frac{ d {\cal R}}{d \left( \frac{\lambda^2}{a} \right)} 
= - \frac{a^2}{\lambda^2} \frac{d{\cal R}}{d a}.
\EE
Note that this is the same transformation rule as for $B$ above.  Thus, the NLS 
scheme-fixing condition, $\frac{d{\cal R}}{da}=B(a)$, transforms into itself.  It is 
straightforward to check that the same is true of the invariants master equation 
Eq.~(\ref{invariantsmastereq}).  
It thus seems that an $a \to {a^*}^2/a$ duality is not special to the BZ limit, but is 
a general property of all-orders NLS and hence of all-orders OPT.

\section{Conclusions}
\setcounter{equation}{0}
\label{ConcSect}

The BZ expansion and RS invariance appear compatible.  While BZ results are most 
simply obtained in a restrictive class of `regular' schemes, the same results emerge 
from `irregular' schemes, though they then require consideration of all orders of 
perturbation theory.  Results in OPT for the fixed-point value ${\cal R}^*$ are never 
far from the BZ result and converge quite nicely to it.  The error at  $\kplus$ order 
shrinks as $\ln^2 k/k^2$ in NLS, as $\ln k/k^2$ in PWMR, and probably slightly 
faster in true OPT.  Our explorations provide some insight into how the subtle features 
of OPT conspire to improve finite-order results.  

It might be claimed that the EC scheme, or any `regular' scheme is clearly {\it better} 
than OPT in the BZ limit, since their results converge {\it immediately} to the right result.  
This is true, but one should keep in mind that the BZ limit, where $n_f$ is infinitesimally 
less than $16\half$, is not a remotely physical theory, even in principle.  It is an open question 
whether or not OPT gives better results than the EC scheme for $n_f=16$, the closest 
physical case.  

The situation with the critical exponent $\gamma^*$ is much less satisfactory.  While 
the all-orders NLS formulas produce the correct result, the finite-order NLS and PWMR 
results do not actually converge.  In true OPT the results might converge but, if so, 
the convergence is extremely slow.  The problem may stem from trying 
to obtain $\gamma^*$ as a by-product of the optimization of ${\cal R}^*$.  If one 
is principally interested in $\gamma^*$ itself, then one should construct its own 
perturbation series and optimize that.  However, our reason here for studying $\gamma^*$ 
was not for its own sake, but as a shortcut to obtaining ${\cal R}(Q)$ at non-zero $Q$, 
relying on Eq.~(\ref{eq: YY}), which holds for the first three orders of the BZ expansion.  
That was very convenient because we only needed the optimization conditions at the 
fixed point, and these are analytically much simpler than for general $Q$.  However, the 
natural procedure is to optimize ${\cal R}(Q)$ itself.  There is no reason to suppose that 
the convergence of OPT for ${\cal R}(Q)$ at non-zero $Q$ is significantly worse than 
for ${\cal R}^*$; indeed, as $Q$ gets larger we expect convergence to become much 
better.  Thus, our difficulties with $\gamma^*$ are probably a technical, mathematical 
issue, rather than a problem of physical concern.

The investigations in this paper have gone off in a number of different directions and 
reveal new territories worthy of further exploration.  A key result is the ``fixed-point 
OPT master equation'' (\ref{fpOPTmastereq}) which opens a route to an analytical treatment 
of arbitrarily high orders of OPT, given knowledge of the $\tilde{\rho}_i$ invariants --- 
although here we have only been able to make progress in two simplifying approximations, 
NLS and PWMR\@. It appears that the simple NLS approximation does yield the all-orders 
limit of OPT, although it is a poor guide to the rate of approach to that limit.  The NLS 
formulas, (\ref{vofu}, \ref{Bofu}) at leading order in the BZ expansion, and 
(\ref{vofu2}, \ref{Bofu2}) at next-to-leading order, are remarkably simple.  They illustrate 
a general $a \to {a^*}^2/a$ duality property of all-orders OPT that is intriguing and 
deserves further study.

We close by mentioning some important developments \cite{Neveu,Siringo} which combine 
RS optimization with the optimization of a variational mass parameter, as in the $\phi^4$ 
anharmonic oscillator problem \cite{CK}-\cite{deltaexp}.  Perhaps the methods discussed 
here can be extended to investigate these approaches at high orders.

\newpage

\section*{Appendix A:  The critical exponent $\boldsymbol{\gamma^*}$}
\renewcommand{\theequation}{A.\arabic{equation}}
\setcounter{equation}{0}

    The critical exponent $\gamma^*$ governing the approach of ${\cal R}$ 
its infrared limit ${\cal R}^*$:
\BE
({\cal R}^*-{\cal R}) \propto Q^{\gamma^*}.
\EE  
is normally thought to be the slope of the $\beta$ function at the fixed 
point \cite{Gross}.  That is not quite true \cite{chylafp}.  The puzzle is 
resolved in Ref.~\cite{effexp}, whose main points we briefly summarize.   

    Since ${\cal R}$ is a physical quantity and $Q$ is a physical parameter, the 
successive logarithmic derivatives of ${\cal R}$:
\BE
{\cal R}_{[n+1]} \equiv Q \frac{d {\cal R}_{[n]}}{dQ}
\EE
for $n=1,2,3,\ldots$, with ${\cal R}_{[1]} \equiv {\cal R}$, must be 
RS-invariant quantities (at any $Q$).  In particular, the combination
\BE
\label{gam1}
\gamma(Q) \equiv \frac{{\cal R}_{[3]}}{{\cal R}_{[2]}} 
\,\, = 1+Q \, \frac{d^2{\cal R}}{dQ^2} \Big/  \frac{d{\cal R}}{dQ}
\EE
is RS invariant.  It is the exponent of the local-power-law form of 
${\cal R}(Q)$ around a specific $Q$.  Standard RG arguments, relating $Q$ 
and $\mu$ dependence, lead to 
\BE
\label{gam2}
\gamma(Q) =   \frac{d \beta}{da} + 
\beta(a) \frac{d^2 {\cal R}}{d a^2} \Big/ \frac{d{\cal R}}{d a} ,
\EE
and one can verify explicitly that this quantity is invariant under RS transformations 
\cite{effexp}.  

   The critical exponent $\gamma^*$ is the infrared-fixed-point limit of 
$\gamma(Q)$.  Since $\beta(a)$ vanishes in this limit one might think that 
the second term in Eq.~(\ref{gam2}) always drops out.  While this is 
often the case, it is not always true, and the NLS, where $d{\cal R}/da$ 
also vanishes at the fixed point, is a case where the second term 
contributes (see Sect.~\ref{BZNLS}).  Quite generally, it is important 
to recognize that $\left. d\beta/da \right|_*$ is {\it not} RS invariant; 
the second term in Eq.~(\ref{gam2}), even though it may vanish in a large 
class of schemes, is crucial to the RS invariance of $\gamma^*$. 

   Another issue arises with finite-order approximations, because 
then the equivalence between Eqs.~(\ref{gam1}) and (\ref{gam2}) is not 
necessarily preserved.  In OPT the two are generally not the same at finite 
$Q$, but, remarkably, they do coincide at $Q=0$ \cite{unfixed}.  We have 
not investigated whether this is also true for NLS and PWMR, which would 
entail explicitly considering ${\cal R}$ at finite $Q$ and then investigating 
its $Q \to 0$ behaviour.

\section*{Appendix B:  Pinch mechanism infrared limit}
\renewcommand{\theequation}{B.\arabic{equation}}
\setcounter{equation}{0}

      As discussed in Ref.~\cite{unfixed}, a finite infrared limit in OPT can occur 
through a pinch mechanism whereby the evolving $B(a)$ function of the 
optimized scheme develops a minimum that ``pinches'' the horizontal axis 
at a ``pinch point'' $\ap$, which ultimately becomes a double zero of $B(a)$.  
The infrared limit of the couplant, however, is at an ``unfixed point'' 
$\adot > \ap$ that is {\it not} a zero of the $\beta$ function.\footnote{
Note the slightly different notation ($\star$ instead of $*$) for infrared-limiting 
quantities according to whether they correspond to an unfixed or a fixed point.
}
The approach to 
the infrared limit is not a power law, but rather \cite{unfixed}
\BE 
{\cal R}^\star - {\cal R} = 
\frac{1}{b_{\rm ir}^2}\frac{1}{\mid\!\ln Q/\tilde{\Lambda}_{{\cal R}}|^2} 
\quad \quad \,\, {\mbox{\rm as}} \,\,\, Q \to 0 ,
\EE
which corresponds to $\gamma^\star=0$ since 
\BE
Q \frac{d {\cal R}}{d Q} \sim - 2 b_{\rm ir} ({\cal R}^\star -{\cal R})^{3/2}
\EE
for ${\cal R}$ close to ${\cal R}^\star$.   In the $k=3$ case, the coefficient 
$b_{\rm ir}$ was found to be 
\BE
b_{\rm ir}^{(k=3)} = \sqrt{2 \ap(3+c \ap)} 
\left(\frac{\ap}{\adot}\right)^2 \frac{b}{\pi} ,
\EE 
and in the $e^+e^-$ case the pinch mechanism was operative for 
$6.7 < n_f < 15.2$.  

     In the BZ limit, $n_f \to 16\half$, the pinch mechanism does not seem to occur 
in true OPT, at least as far as we have been able to explore it in Sect.~\ref{Loword}.  
However, the mechanism is probably close to being relevant because in the BZ limit the 
critical exponent $\gamma^* \sim b a_0$ tends to zero.  A small or zero $\gamma^*$ 
gives rise to a sharp infrared ``spike'' in ${\cal R}$ plotted versus $Q$, as in Fig.~1.

   The NLS and PWMR approximations to OPT seem to have fixed points only in every 
other order (for odd $k$ in NLS, and even $k$ in PWMR).   In these orders, as discussed 
in Sect.~\ref{BZNLS}, the $B(u)$ function closely approximates its limiting form 
$(1-u)/(1+u)^3$ until $u$ gets close to $1$, when it suddenly dives down, 
producing a zero.  In the alternating orders $B(u)$ suddenly shoots upwards and there 
is no zero.  However, $B(u)$ then has a minimum very close to the horizontal axis, so only 
a slight modification of the scheme would produce a ``pinch point.'' 
 
   We first show that that, in circumstances where the pinch mechanism {\it does} govern the 
infrared limit of OPT, the master equation that replaces Eq.~(\ref{fpOPTmastereq}) is              
\BE
\label{unfixedOPTmastereq}
\frac{d{\cal R}}{da}= \left( \frac{1-a/\adot}{1-a/\ap} \right) 
\left[
B(a)- \frac{a}{(k-1)} \left( 2 \frac{d B(a)}{da} + \frac{B(a)}{(a_p-a)} \right) 
\right] .
\EE
(Superscripts ``${\scriptstyle{(k+1)}}$'' on ${\cal R}$ and $B(a)$ are omitted for brevity.) 
Except for the pre-factor, and the fact that $\ap$ (not $\adot$) replaces $a^*$ in the 
last term, this equation is identical to (\ref{fpOPTmastereq}).   

    The derivation is as follows.  As $Q \to 0$ the $B(a)$ function nearly vanishes at 
the pinch point $\ap$ and close to $\ap$ can be approximated by the form \cite{unfixed} 
\BE
\label{Bform}
B(a) \approx \eta \left( (a-\ap)^2 + \delta^2 \right),
\EE
where $\delta$ vanishes $\propto 1/\!\mid\! \ln Q \!\mid$ as $Q \to 0$ and $\eta$ 
is some positive constant.  The integrals $I_j(a)$ of Eq.~(\ref{Ij}) are dominated by a 
huge peak in their integrands around $\ap$:
\BE
\label{Ijform}
I_j(a) \approx \int \! dx 
\frac{x^{j-2}}{\left( \eta \left( (a-\ap)^2 + \delta^2 \right) \right)^2} 
\approx \frac{\ap^{j-2}}{\eta^2} \frac{\pi}{2 \delta^3}.
\EE
One can thus obtain the $\delta \to 0$ behaviour of the $B_j(a)$ and 
hence the $H_j$ functions \cite{unfixed}.  (Note that the $B(a)/a^{j-1}$ factor in 
Eq.~(\ref{Bj}) will involve the limiting value of $a$, which is $\adot$ and 
not $\ap$.)  While the $B_j$'s and $H_j$'s diverge, the $1/\delta^3$ 
factors cancel out, as does $\eta$, in Eq.~(\ref{formula}), leaving finite 
limiting values for the optimized $r_m$ coefficients.  Instead of 
Eq.~(\ref{fpOPTcond}) of the fixed-point case, we find 
\BE
s_m {\adot}^m = \frac{1}{(k-1)} \left[ 
\left(\frac{\adot}{\ap}\right)^m \sum_{j=0}^{m} (k-m-j-1) c_j \ap^j 
- \left(\frac{\adot}{\ap}\right)^{m-1} \sum_{j=0}^{m-1} (k-m-j) c_j \ap^j 
\right] ,
\EE
where $s_m \equiv (m+1) r_m$.  Using a dummy variable $a$ we can then form 
the function
\BE 
{\cal S}(a) = \frac{d{\cal R}}{da} = \sum_{m=0}^{k} s_m a^m .
\EE
Reorganizing the resulting double summation over $m$ and $j$ so that the latter 
becomes the outer summation, the inner summations become finite geometric 
series or derivatives thereof.  The outer $j$ summation then produces terms 
that are $B(a)$ or $dB/da$ or $B(\ap)$ or $ \left. dB/da \right|_{a=\ap}$.  The 
last two vanish in the infrared limit since $\ap$ is then a double zero of the 
$B(a)$ function.  After some further algebraic tidying up the result reduces to 
Eq.~(\ref{unfixedOPTmastereq}) above.

     Note that the na\"{\i}ve large-$k$ limit of Eq.~(\ref{unfixedOPTmastereq}) 
is not the NLS condition (\ref{NLSmastereq}) but 
\BE 
\label{unNLSmastereq}
\frac{d{\cal R}}{da}= \left( \frac{1-a/\adot}{1-a/\ap} \right) B(a)  
\quad\quad\quad {\mbox{\rm (NLS}}^\prime).
\EE
If we proceed in parallel with the analysis in Sect.~\ref{BZNLS} we find, instead 
of Eq.~(\ref{NLSvu}),
\BE
\int \frac{dv}{v \sqrt{1-v}} = \int \frac{du}{u} 
\sqrt{ \frac{1-u/\udot}{1-u/\up} } .
\EE
Note that the above equations correspond to the {\it ansatz} form used in the PWMR 
analysis of Sect.~\ref{PWMRSect} with $\xi$ replaced by
\BE
\xi \rightarrow \sqrt{ \frac{1-u/\udot}{1-u/\up} }.
\EE
Doing the integrations, exponentiating both sides, and solving for $v$ leads to 
\BE
v = \frac{4 U}{(1+U)^2},
\EE
where
\BE
U= \left( \frac{4 \udot \up}{\udot-\up} \right) \left(
\frac{ \sqrt{\frac{1-u/\udot}{1-u/\up}} -1 }{ \sqrt{\frac{1-u/\udot}{1-u/\up}} +1 } 
\right) 
\left( \frac{ \sqrt{\udot-u} + \sqrt{\up-u} }{\sqrt{\udot} + \sqrt{\up}} 
\right)^{ 2 \sqrt{ \frac{\up}{\udot} } } .
\EE
Note that when $u>\up$ (which is relevant since $u$ ranges from $0$ to $\udot$, which 
must exceed $\up$) this formula for $U$ develops an imaginary part.  
However, recall that both $v$ and $B$, 
\BE
\label{}
B = \frac{(1-U)}{(1+U)^3} \frac{U}{u}  \sqrt{ \frac{1-u/\up}{1-u/\udot} }   
\EE
(Cf. Eq.~(\ref{Bfromxi})), have to be expanded as series in $u$ and then 
{\it truncated} after $k$ terms, making them inevitably real.  

     These formulas are hard to deal with, even at low orders, especially since 
$\up$ and $\udot$ have to be determined by the requirements that the truncated 
$B$ and its derivative vanish at the pinch point $\up$.  For $k=2, 4$ there does not 
seem to be any viable solution, but for sufficiently large $k$ it appears there is.  
Anticipating that both $\up$ and $\udot$ will tend to $1$ as $k \to \infty$, we define
\BE 
\delta \equiv \frac{1}{\up} - \frac{1}{\udot} 
\EE
and proceed to expand to lowest non-trivial order in $\delta$.  This gives
\BE
U \approx u \left( 1- \frac{\delta}{2} \ln(1-u) \right),
\EE
\BE
v \approx \frac{4u}{(1+u)^2} - 2 \delta u \frac{(1-u)}{(1+u)^3} \ln(1-u),
\EE
and 
\BE
B \approx \frac{1-u}{(1+u)^3} -\frac{\delta}{2} \left( 
\frac{u}{(1+u)^3} + \frac{(1-4u+u^2)}{(1+u)^4} \ln(1-u) \right) .
\EE
Remarkably, one can find analytic expressions for the truncated-series versions 
of $v$ and $B$ and thereby explore numerical results up to very high $k$ values.  
These results (see table \ref{Pinchtable}) show that indeed there a valid solution (with $\udot >\up$) 
exists with $\delta$ tending to zero as $\delta \sim (2/\ln2)(1/k)$ and 
${\cal R}^\star/a_0$ tending to $1$.  

\begin{table}[htbp]
\begin{center}
\begin{tabular}[b]{|r||d{5}|d{5}|d{5}|d{5}|}
\hline
$k$ \quad & 
\multicolumn{1}{c|}{$\, \up \,\,$} & 
\multicolumn{1}{c|}{$\, \udot \,\,$} & 
\multicolumn{1}{c|}{$\delta$}  &
\multicolumn{1}{c|}{$v^\star = \frac{{\cal R}^\star}{a_0}$}  \\
\hline
$100$          &  0.95018      &  0.97735      &  0.02925  & 0.71485        \\
$600$          &  0.98819      &  0.99292      &  0.00482  & 0.95982        \\
$10,000$      &  0.99895      &  0.99924      &  0.00029   & 0.99856        \\
\hline

\end{tabular}
\end{center}
\vspace*{-5mm}
\caption{ {\textsl{NLS$^\prime$ results, to lowest-order in $\delta$, in the BZ limit
}}}

\label{Pinchtable}
\end{table}


\newpage
                                                                                                                                                                                                                                                                                                                                                                                                                                                                                                                                  
\begin{thebibliography} {99}

\bibitem{BZ}
  T. Banks and A. Zaks, Nucl. Phys. B {\bf 196}, 189 (1982).

\bibitem{White}
  A. R. White, Phys. Rev. D {\bf 29}, 1435 (1984); Int. J. Mod. Phys. A 
  {\bf 8}, 4755 (1993).

\bibitem{grun}
  G. Grunberg, Phys. Rev. D {\bf 46}, 2228 (1992).

\bibitem{BZlett}
  P. M. Stevenson, Phys. Lett. B {\bf 331}, 187 (1994).

\bibitem{RG} 
  E. C. G. Stueckelberg and A. Peterman, Helv. Phys. Acta {\bf 26}, 449 (1953); 
  M. Gell Mann and F. Low, Phys. Rev. {\bf 95}, 1300 (1954); 
  N. N. Bogoliubov and D. V. Shirkov, {\it Introduction to the theory of quantized fields} 
  (Interscience, New York, 1959).

\bibitem{OPT} 
  P. M. Stevenson, Phys. Rev. D {\bf 23}, 2916 (1981).

\bibitem{Caveny}
  S. Caveny and P.~M.~Stevenson, hep-ph/9705319 (unpublished) Appendix B.   

\bibitem{KSS} 
  J. Kubo, S. Sakakibara, and P. M. Stevenson, Phys. Rev. D {\bf 29}, 1682 (1984).

\bibitem{CKL}
  J. Ch\'yla, A. Kataev, and S. A. Larin, Phys. Lett. B {\bf 267}, 269 (1991).

\bibitem{lowen}  
  A. C. Mattingly and P. M. Stevenson, Phys. Rev. Lett. {\bf 69}, 1320 (1992); 
  Phys. Rev. D {\bf 49}, 437 (1994).

\bibitem{GardiK}
   E. Gardi and M. Karliner, Nucl. Phys. B {\bf 529}, 383 (1998).

\bibitem{OPTnew}
  P. M. Stevenson, Nucl. Phys. B {\bf 868}, 38 (2013).

\bibitem{Posolda}
  P. Posolda, J. Phys. G {\bf 41} 095007 (2014).

\bibitem{unfixed}
  P. M. Stevenson, Nucl. Phys. B {\bf 875}, 63 (2013).  

\bibitem{optult} 
  P. M. Stevenson, Nucl. Phys. B {\bf 231}, 65 (1984).

\bibitem{ZinnJ}
  R. Seznec and J. Zinn-Justin, J. Math. Phys. {\bf 20}, 1398 (1979). 

\bibitem{IndConvAHO}
  I. R. C. Buckley, A. Duncan, and H. F. Jones, Phys. Rev. D {\bf 47}, 2554 (1993); 
  A. Duncan and H. F. Jones, Phys. Rev. D {\bf 47}, 2560 (1993); 
  C. M. Bender, A. Duncan, and H. F. Jones, Phys. Rev. D {\bf 49}, 4219 (1994); 
  C. Arvanitis, H. F. Jones, and C. S. Parker, Phys. Rev. D {\bf 52}, 3704 (1995);
  R. Guida, K. Konishi, and H. Suzuki, Ann. Phys. {\bf 241}, 152 (1995);
  W. Janke and H. Kleinert, Phys. Rev. Lett. {\bf 75}, 2787 (1995); 
  H. Kleinert and V. Schulte-Frohlinde, {\it Critical properties of $\phi^4$ theories}, 
  (World Scientific, Singapore, 2001), Chap. 19.  

\bibitem{CK}
  W. Caswell,  Ann. Phys. (N.Y.) {\bf 123}, 153 (1979); J. Killingbeck, J. Phys. A 
  {\bf 14}, 1005 (1981); E. J. Austin and J. Killingbeck, {\it ibid} {\bf 15}, 
  L443 (1982).
 
 \bibitem{GEP}
  T. Barnes and G. I. Ghandour, Phys. Rev. D {\bf 22}, 924 (1980);
  P. M. Stevenson, Phys. Rev. D {\bf 30}, 1712 (1984); 
  {\it ibid.} {\bf 32}, 1389 (1985); 
  I. Stancu and P. M. Stevenson, {\it ibid.} {\bf 42}, 2710 (1990). 

\bibitem{deltaexp}
  A. Duncan and M. Moshe, Phys. Lett. B {\bf 215}, 352 (1988); 
  H. F. Jones, Nucl. Phys. B (Proc. Suppl.) {\bf 16}, 592 (1990);
  A. Duncan and H. F. Jones, Phys. Rev. D {\bf 47}, 2560 (1993); 
  M. Pinto and R. Ramos, Phys. Rev. D {\bf 60}, 105005 (1999).
 
\bibitem{Acoleyen}
  K. Van Acoleyen and H. Verschelde, Phys. Rev. D {\bf 69} 125006 (2004); 
  K.~Van~Acoleyen, Ph.D. thesis, University of Ghent (2002) (unpublished);     
  J.~Fraser, Masters Thesis, University of Durham (2012) (unpublished).

\bibitem{Beneke}
  M. Beneke, Nucl. Phys. B {\bf 405}, 424 (1993).

\bibitem{Gross}
  D. J. Gross, in {\it Methods in Field Theory}, edited by R.~Balian and 
  J.~Zinn-Justin (North-Holland, Amsterdam, 1976);
  A.~Peterman, Phys. Reports {\bf 53C}(3), 157 (1979).

\bibitem{chylafp}
  J. Ch\'{y}la, Phys. Rev. D {\bf 38}, 3845 (1988).

\bibitem{effexp}
 P. M. Stevenson, arXiv: 1606.06951 [hep-ph].

\bibitem{CG} 
  W. Celmaster and R. J. Gonsalves, Phys. Rev. D {\bf 20}, 1420 (1979). 

\bibitem{Grunberg}
  G. Grunberg, Phys. Rev. D {\bf 29}, 2315 (1984); A. Dhar and V. Gupta, 
  Phys. Rev. D {\bf 29}, 2822 (1984); 

\bibitem{PWMR} 
  M. R. Pennington, Phys. Rev. D {\bf 26}, 2048 (1982); 
  J. C. Wrigley, Phys. Rev. D {\bf 27}, 1965 (1983);  
  See also P. M. Stevenson, Phys. Rev. D, {\bf 27}, 1968 (1983); 
  J. A. Mignaco and I. Roditi, Phys. Lett. B {\bf 126}, 481 (1983).

\bibitem{Watson}
  G. N. Watson, Trans. Cam. Phil. Soc. {\bf 22}, 277 (1918); 
  Y. L. Luke, {\it The special functions and their approximations}, Vol. 1, (Academic Press, 
  New York, 1969).

\bibitem{Neveu}
  J.-L. Kneur and A. Neveu, Phys. Rev. D {\bf 81}, 125012 (2010);
  {\it ibid.} {\bf 85}, 014005 (2012); 
  {\it ibid.} {\bf 88}, 0704025 (2013);
  {\it ibid.} {\bf 92}, 074027 (2015).

\bibitem{Siringo}
  F. Siringo, Nucl. Phys. B {\bf 907}, 572 (2016); 
  hep-ph 1509.05891; hep-ph 1605.07357.


  
\end{thebibliography}
\end{document}